\RequirePackage[l2tabu, orthodox]{nag}
\documentclass{sig-alternate}

\usepackage{xcolor}
\usepackage{graphicx}
\usepackage{amsmath}
\usepackage{amssymb}
\usepackage{color}
\usepackage{ifpdf}
\usepackage[switch,columnwise]{lineno}

\usepackage{float}
\usepackage{srcltx}
\usepackage[utf8]{inputenc}
\usepackage{multirow}
\usepackage{rotating}
\usepackage{subfigure}

\usepackage{url}
\usepackage{booktabs}
\usepackage{listings}   
\usepackage{paralist}    
\usepackage{wrapfig}    
\usepackage{multirow}
\usepackage{ifpdf}
\usepackage{xspace}
\usepackage{keyval}  
\usepackage{color}
\usepackage{comment}
\usepackage{pdfsync}

\definecolor{listinggray}{gray}{0.95}
\definecolor{darkgray}{gray}{0.7}
\definecolor{commentgreen}{rgb}{0, 0.4, 0}
\definecolor{darkblue}{rgb}{0, 0, 0.4}
\definecolor{middleblue}{rgb}{0, 0, 0.7}
\definecolor{darkred}{rgb}{0.4, 0, 0}
\definecolor{brown}{rgb}{0.5, 0.5, 0}

\usepackage[normalem]{ulem}
\makeatletter
\def\cyanuwave{\bgroup \markoverwith{\lower3.5\p@\hbox{\sixly \textcolor{cyan}{\char58}}}\ULon}
\def\reduwave{\bgroup \markoverwith{\lower3.5\p@\hbox{\sixly \textcolor{red}{\char58}}}\ULon}
\def\blueuwave{\bgroup \markoverwith{\lower3.5\p@\hbox{\sixly \textcolor{blue}{\char58}}}\ULon}
\font\sixly=lasy6 
\makeatother

\newif\ifdraft
\ifdraft

\newcommand{\onote}[1]{ {\textcolor{magenta} { ***Ole: #1 }}}
\newcommand{\terminology}[1]{ {\textcolor{red} {(Terminology used: \textbf{#1}) }}}

\newcommand{\jhanote}[1]{ {\textcolor{red} { ***shantenu: #1 }}}
\newcommand{\alnote}[1]{ {\textcolor{green} { ***andreL: #1 }}}
\newcommand{\amnote}[1]{ {\textcolor{blue} { ***andreM: #1 }}}
\newcommand{\smnote}[1]{ {\textcolor{brown} { ***sharath: #1 }}}
\newcommand{\pmnote}[1]{ {\textcolor{brown} { ***Pradeep: #1 }}}
\newcommand{\msnote}[1]{ {\textcolor{cyan} { ***mark: #1 }}}
\newcommand{\mrnote}[1]{ {\textcolor{purple} { ***melissa: #1 }}}
\newcommand{\aznote}[1]{ {\textcolor{orange} { ***ashley: #1 }}}
\newcommand{\mtnote}[1]{ {\textcolor{orange} { ***matteo: #1 }}}
\newcommand{\note}[1]{ {\textcolor{magenta} { ***Note: #1 }}}
\else
\newcommand{\onote}[1]{}
\newcommand{\terminology}[1]{}

\newcommand{\alnote}[1]{}
\newcommand{\amnote}[1]{}
\newcommand{\aznote}[1]{}
\newcommand{\athotanote}[1]{}
\newcommand{\smnote}[1]{}
\newcommand{\pmnote}[1]{}
\newcommand{\jhanote}[1]{}
\newcommand{\msnote}[1]{}
\newcommand{\mtnote}[1]{}
\newcommand{\note}[1]{}
\newcommand{\mrnote}[1]{}
\fi

\newcommand{\pilot}{Pilot\xspace}
\newcommand{\pilots}{Pilots\xspace}
\newcommand{\pilotjob}{Pilot-Job\xspace}
\newcommand{\pilotjobs}{Pilot-Jobs\xspace}

\newcommand{\pilotdata}{Pilot-Data\xspace}

\newcommand{\MW}{master-worker\xspace}
\newcommand{\panda}{PanDA\xspace}

\lstdefinestyle{myListing}{
  frame=single,   
  backgroundcolor=\color{listinggray},  
  language=C,       
  basicstyle=\ttfamily \footnotesize,
  breakautoindent=true,
  breaklines=true
  tabsize=2,
  captionpos=b,  
  aboveskip=0em,
  belowskip=-2em,
}      

\lstdefinestyle{myPythonListing}{
  frame=single,   
  backgroundcolor=\color{listinggray},  
  language=Python,       
  basicstyle=\ttfamily \footnotesize,
  breakautoindent=true,
  breaklines=true
  tabsize=2,
  captionpos=b,  
}

\ifpdf
\DeclareGraphicsExtensions{.pdf, .jpg, .tif}
\else
\DeclareGraphicsExtensions{.eps, .jpg, .ps}
\fi

\usepackage{cite}
\usepackage{paralist}
\usepackage{multirow}
\usepackage[activate={true,nocompatibility},final,tracking=true,kerning=true,spacing=true]{microtype}
\usepackage{hyperref}

\microtypecontext{spacing=nonfrench}

\begin{document}
\CopyrightYear{2015}

\title{A Comprehensive Perspective on \pilotjob Systems}


\numberofauthors{3}

\author{
  \alignauthor Matteo Turilli \\
    \affaddr{RADICAL Laboratory, ECE}\\
    \affaddr{Rutgers University}\\
    \affaddr{New Brunswick, NJ, USA}\\
    \email{matteo.turilli@rutgers.edu}
  \alignauthor Mark Santcroos\\
    \affaddr{RADICAL Laboratory, ECE}\\
    \affaddr{Rutgers University}\\
    \affaddr{New Brunswick, NJ, USA}\\
    \email{mark.santcroos@rutgers.edu}
  \and
  \alignauthor Shantenu Jha\titlenote{Corresponding author.}\\
    \affaddr{RADICAL Laboratory, ECE}\\
    \affaddr{Rutgers University}\\
    \affaddr{New Brunswick, NJ, USA}\\
    \email{shantenu.jha@rutgers.edu}
}

\maketitle

\begin{abstract}

  \pilotjob systems play an important role in supporting distributed scientific
  computing. They are used to consume more than 700 million CPU hours a year by
  the Open Science Grid communities, and by processing up to 1 million jobs a
  day for the ATLAS experiment on the Worldwide LHC Computing Grid. With the
  increasing importance of task-level parallelism in high-performance
  computing,  \pilotjob systems are also witnessing an adoption beyond
  traditional domains. Notwithstanding the growing impact on scientific
  research, there is no agreement upon a definition of \pilotjob system and no
  clear understanding of the underlying abstraction and paradigm. \pilotjob
  implementations have proliferated with no shared best practices or open
  interfaces and little interoperability. Ultimately, this is hindering the
  realization of the full impact of \pilotjobs by limiting their robustness,
  portability, and maintainability. This paper offers a comprehensive analysis
  of \pilotjob systems critically assessing their motivations, evolution,
  properties, and implementation. The three main contributions of this paper
  are: (i) an analysis of the motivations and evolution of \pilotjob systems;
  (ii) an outline of the \pilot abstraction, its distinguishing logical
  components and functionalities, its terminology, and its architecture
  pattern; and (iii) the description of core and auxiliary properties of
  \pilotjobs systems and the analysis of seven exemplar \pilotjob
  implementations. Together, these contributions illustrate the \pilot
  paradigm, its generality, and how it helps to address some challenges in
  distributed scientific computing.

\end{abstract}

\keywords{\pilotjob, distributed applications, distributed systems.}

%
\section{Introduction}
\label{sec:intro}

\pilotjobs provide a multi-stage mechanism to execute workloads. Resources are
acquired via a placeholder job and subsequently assigned to workloads.
\pilotjobs are having a high impact on scientific and distributed
computing~\cite{katz2012}. They are used to consume more than 700 million CPU
hours a year~\cite{osg_presentation_url} by the Open Science Grid
(OSG)~\cite{pordes2007,osg_url} communities, and process up to 1 million jobs a
day~\cite{de2014panda} for the ATLAS experiment~\cite{aad2008atlas} on theLarge
Hadron Collider (LHC)~\cite{lhc1995large} Computing Grid
(WLCG)~\cite{bonacorsi2007wlcg,wlcg_url}. A variety of \pilotjob systems are
used on distributed computing infrastructures (DCI):
Glidein/GlideinWMS~\cite{glidein_presentation_url,sfiligoi2008glideinwms}, the
Coaster System~\cite{hategan2011coasters}, DIANE~\cite{moscicki2003diane},
DIRAC~\cite{casajus2010dirac}, \panda~\cite{chiu2010pilot},
GWPilot~\cite{rubio2015gwpilot}, Nimrod/G~\cite{buyya2000nimrod},
Falkon~\cite{raicu2007falkon}, MyCluster~\cite{walker2006creating} to name a
few.


A reason for the success and proliferation of \pilotjob systems is that they
provide a simple solution to the rigid resource management model historically
found in high-performance and distributed computing. \pilotjobs break free of
this model in two ways: (i) by using late binding to make the selection of
resources easier and more
effective~\cite{moscicki2011,glatard2010,delgado2014}; and (ii) by decoupling
the workload specification from the management of its execution. Late binding
results in the ability to utilize resources dynamically, i.e., the workload is
distributed onto resources only when they are effectively available. Decoupling
workload specification and execution simplifies the scheduling of workloads on
those resources.

In spite of the success and impact of \pilotjobs, we perceive a problem: the
development of \pilotjob systems has not been grounded on an analytical
understanding of underpinning abstractions, architectural patterns, or
computational paradigms.  The properties and functionalities of \pilotjobs have
been understood mostly, if not exclusively, in relation to the needs of the
containing software systems or on use cases justifying their immediate
development.




 
These limitations have also resulted in a fragmented software landscape, where
many \pilotjob systems lack generality, interoperability, and robust
implementations.  This has led to a proliferation of functionally equivalent
systems motivated by similar objectives that often serve particular use cases
and target particular resources.


Addressing the limitations of \pilot systems while improving our general
understanding of \pilotjob systems is a priority due to the role they will play
in the next generation of high-performance computing. Most existing
high-performance system software and middleware are designed to support the
execution and optimization of single tasks. Based on their current utilization,
\pilotjobs have the potential to support the growing need for scalable
task-level parallelism and dynamic resource management in high-performance
computing~\cite{raicu2007falkon, hategan2011coasters, review_radicalpilot_2015}.

The causes of the current status quo of \pilotjob systems are social, economic,
and technical. While social and economic considerations may play a determining
role in promoting fragmented solutions, this paper focuses on the technical
aspects of \pilotjobs. We contribute a critical analysis of the current state
of the art describing the technical motivations and evolution of \pilotjob
systems, their characterizing abstraction (the \pilot abstraction), and the
properties of their most representative and prominent implementations. 
Our analysis will yield the \pilot paradigm, i.e., the way in which
\pilotjobs are used to support and perform distributed computing.

The remainder of this paper is divided into four sections. \S\ref{sec:history}
offers a description of the technical motivations of \pilotjob systems and of
their evolution.

In~\S\ref{sec:understanding}, the logical components and functionalities
constituting the \pilot abstraction are discussed.  We offer a terminology
consistent across \pilotjob implementations, and an architecture pattern for
\pilotjobs systems is derived and described.

In~\S\ref{sec:analysis}, the focus moves to \pilotjob implementations and to
their core and auxiliary properties. These properties are described and then
used alongside the \pilot abstraction and the pilot architecture pattern to
describe and compare exemplar \pilotjob implementations.

In~\S\ref{sec:discussion}, we outline the \pilot paradigm, arguing for its
generality, and elaborating on how it impacts and relates to both other
middleware and applications. Insight is offered about the future directions and
challenges faced by the \pilot paradigm and its \pilotjob systems.

%
\section{Evolution of Pilot-Job Systems}
\label{sec:history}

Three aspects of \pilotjobs are investigated in this paper: the \pilotjob
system, the \pilotjob abstraction, and the \pilotjob paradigm. A \pilotjob
system is a type of software, the \pilotjob abstraction is the set of
properties of that type of software, and the \pilotjob paradigm is the way in
which \pilotjob systems enable the execution of workloads on resources. For
example, DIANE is an implementation of a \pilotjob system; its components and
functionalities are elements of the \pilotjob abstraction; and the type of
workloads, the type of resources, and the way in which DIANE executes the
former on the latter are features of the \pilotjob paradigm.

This section introduces \pilotjob systems by investigating their technical
origins and motivations alongside the chronology of their development.

\subsection{Technical Origins and Motivations}
\label{sec:histabstr}

Five features need elucidation to understand the technical origins and
motivations of \pilotjob systems: task-level distribution and parallelism,
\MW pattern, multi-tenancy, multi-level scheduling, and resource placeholding.
\pilotjob systems coherently integrate resource placeholders, multi-level
scheduling, and coordination patterns to enable task-level distribution and
parallelism on multi-tenant resources. The analysis of each feature clarifies
how \pilotjob systems support the execution of workloads comprised of multiple
tasks on one or more distributed machine.

\textbf{Task-level distribution and parallelism} on multiple resources can be
traced back to 1922 as a way to reduce the time to solution of differential
equations~\cite{richardson1922weather}. In his Weather Forecast
Factory~\cite{lynch1999richardson}, Lewis Fry Richardson imagined distributing
computing tasks across 64,000 ``human computers'' to be processed in parallel.
Richardson's goal was exploiting the parallelism of multiple processors to
reduce the time needed for the computation. Today, task-level parallelism is
commonly adopted in weather forecasting on modern high performance
machines\footnote{A high-performance machine indicates a cluster of computers
  delivering higher performances than single workstations or desktop computers,
  or a resource with adequate performance to support multiple science and
  engineering applications concurrently.} as computers. Task-level parallelism
is also pervasive in computational science~\cite{wegner1997interaction} (see
Ref.~\cite{enmd} and references therein).

\textbf{Master-worker} is a coordination pattern commonly used for distributed
computations~\cite{goux2000enabling,heymann2000adaptive,carriero1995adaptive,sarmenta1999bayanihan,schmidt2002tree}. Submitting tasks to multiple computers
at the same time requires coordinating the process of sending and receiving
tasks; of executing them; and of retrieving and aggregating their
outputs~\cite{coulouris2005distributed}. In the \MW pattern, a ``master'' has a
global view of the overall computation and of its progress towards a solution.
The master distributes tasks to multiple ``workers'', and retrieves and
aggregates the results of each worker's computation. Alternative coordination
patterns have been devised, depending on the characteristics of the computed
tasks but also on how the system implementing task-level distribution and
parallelism has been designed~\cite{freisleben1997coordination}.

\textbf{Multi-tenancy} defines how high-performance machines are exposed to
their users. Job schedulers, often called ``batch queuing
systems''~\cite{czajkowski1998} and first used in the time of punched
cards~\cite{katz1966,silberschatz1998operating}, adopt the batch processing
concept to promote efficient and fair resource sharing. Job schedulers enable
users to submit computational tasks called ``jobs'' to a queue. The execution
of these jobs is delayed waiting for the required amount of the machine's
resources to be available. The extent of delay depends on the number, size, and
duration of the submitted jobs, resource availability, and policies (e.g., fair
usage).

The resource provisioning of high-performance machines is limited, irregular,
and largely unpredictable~\cite{downey1997,wolski2003,li2004,tsafrir2007}. By
definition, the resources accessible and available at any given time can be
fewer than those demanded by all the active users. The resource usage patterns
are also not stable over time and alternating phases of resource availability
and starvation are common~\cite{Furlani2013,Lu2013}. This landscape has
promoted continuous optimization of the resource management and the development
of alternative strategies to expose and serve resources to the users.

\textbf{Multi-level scheduling} is one of the strategies used to improve
resource access across high-performance machines. In multi-level scheduling, a
global scheduling decision results from a set of local scheduling
decisions~\cite{li2008multi,weikum1985theoretical}. For example, an application
submits tasks to a scheduler that schedules those tasks on the schedulers of
individual high-performance machines. While this approach can increase the
scale of applications, it also introduces complexities across resources,
middleware, and applications.

Several approaches have been devised to manage these complexities~\cite{singh2005,ramakrishnan2006toward,foster2008,juve2008,villegas2012,song2009,taylor2007workflows,curcin2008scientific,balderrama2012scalable}
but one of the persistent issues is the increase of the implementation burden
imposed on applications. For example, in spite of progress made by grid
computing~\cite{berman2003grid,foster2003grid} to transparently integrate
diverse resources, most of the requirements involving the coordination of task
execution still reside with the
applications~\cite{legrand2003,krauter2002,darema2005}. This translates into
single-point solutions, extensive redesign and redevelopment of existing
applications when adapted to new use cases or new high-performance machines,
and lack of portability and interoperability.

\textbf{Resource placeholders} are used as a pragmatic solution to better
manage the complexity of executing applications. A resource placeholder
decouples the acquisition of compute resources from their use to execute the
tasks of an application. For example, resources are acquired by scheduling a
job onto a high-performance machine which, when executed, is capable of
retrieving and executing application tasks itself.

Resource placeholders bring together mul\-ti-\-le\-vel sche\-du\-ling and
task-level distribution and parallelism. Placeholders are scheduled on one or
more machines and then multiple tasks are scheduled at the same time on those
placeholders. Tasks can then be executed concurrently and in parallel when the
placeholders covers multiple compute resources. The \MW pattern is often an
effective choice to manage the coordination of tasks execution.

It should be noted that resource placeholders also mitigate the side-effects of
multi-tenancy. A placeholder still spends a variable amount of time waiting to
be executed on a high-performance machine, but, once executed, the application
exerts total control over the placeholder resources. In this way, tasks are
directly scheduled on the placeholder without competing with other users for
the same resources.

Resource placeholders are programs with specific queuing and scheduling
capabilities. They rely on jobs submitted to a high-performance machine to
execute a program with diverse capabilities. For example, jobs usually execute
non interactive programs, but users can submit jobs that execute terminals,
debuggers, or other interactive software.

\subsection{Chronological Evolution}
\label{sec:histimpl}

\begin{figure}[t]
  \centering
    \includegraphics[width=0.48\textwidth]{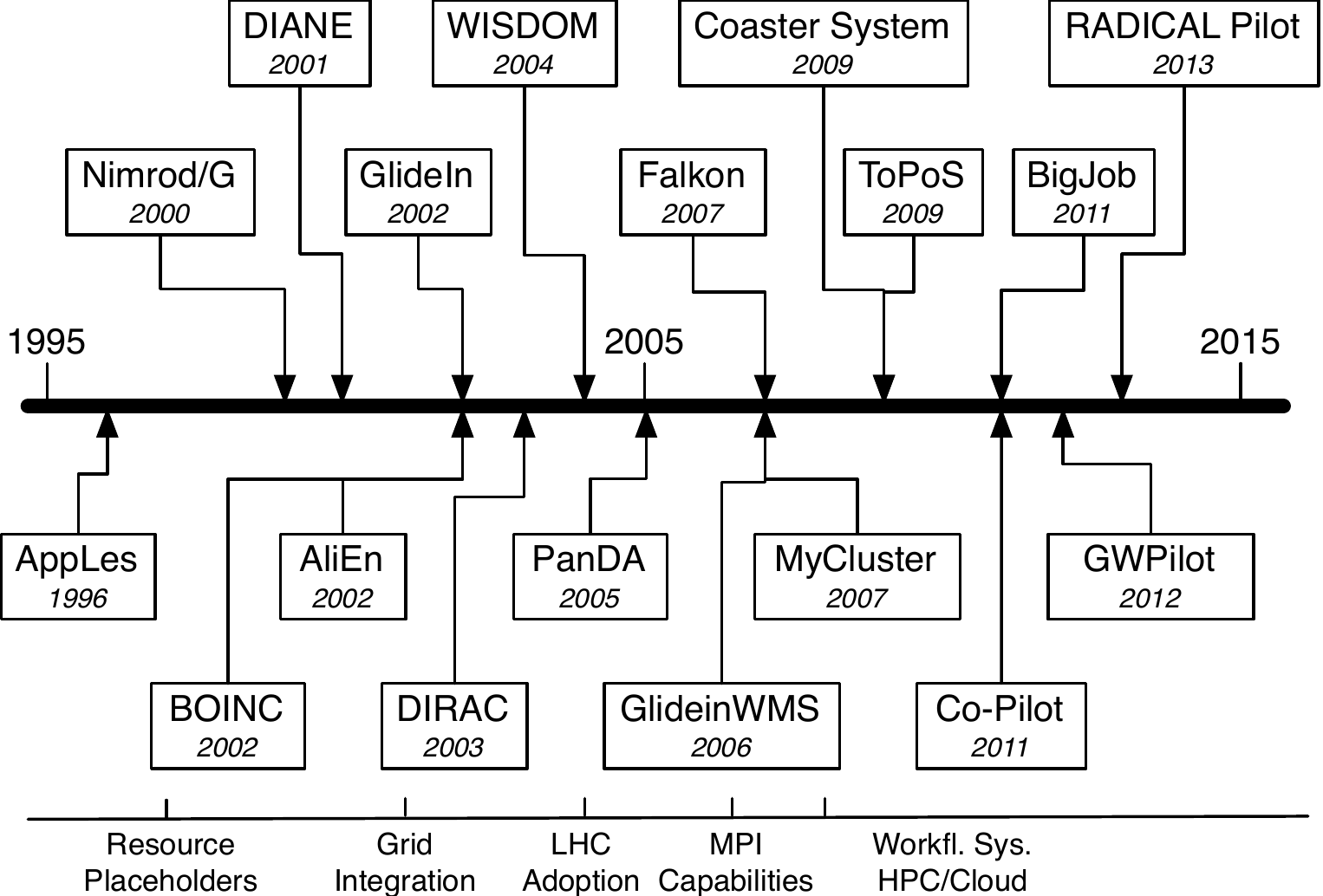}
    \caption{Introduction of \pilotjob systems over time alongside some
      exemplar milestones of their evolution. When available, the date of first
      mention in a publication or otherwise the release date of software
      implementation is used.}
    \label{fig:timeline}
\end{figure}

Figure~\ref{fig:timeline} shows the introduction of \pilotjob systems over time
alongside some of the defining milestones of their evolution.\footnote{To the
best of the authors' knowledge, the term ``pilot'' was first coined in 2004 in
the context of the WLCG Data Challenge\cite{wlcg_url,bonacorsi2007wlcg}, and
then introduced in writing as ``pilot-agent'' in a 2005 LHCb
report~\cite{nobrega2005lhcb}.} This is an approximated chronology based on the
date of the first publication, or when publications are not available, on the
date of the systems' code repository.

The evolution of \pilotjob systems began with the implementation of resource
placeholders to explore application-side task scheduling and high-throughput
task execution. Prototypes of \pilotjob systems followed, eventually evolving
into production-grade systems supporting specific types of applications and
high-performance machines. Recently, \pilot systems have been employed to
support a wide range of workloads and applications (e.g., MPI, data-driven
workflows, tightly and loosely coupled ensembles), and more diverse
high-performance machines (e.g., MPI, data-driven workflows, tightly and
loosely coupled ensembles).


AppLeS (Application Level Schedulers)~\cite{berman1996application} offered an
early implementation of resource placeholders. Developed around 1997, AppLeS
provided an agent that could be embedded into an application to acquire
resources and to schedule tasks onto them. AppLeS provided application-level
scheduling but did not isolate the application from resource acquisition. Any
change in the agent directly translated into a change of the application code.
AppLeS Templates~\cite{berman2003adaptive} was developed to address this issue,
each template representing a class of applications (e.g., parameter
sweep~\cite{casanova2000apples}) that could be adapted to the requirements of a
specific realization.

Volunteer computing projects started around the same time as AppLeS was
introduced. In 1997, the Great Internet Mersenne Prime Search
effort~\cite{woltman2004great}, shortly followed by
distributed.net~\cite{lawton2000distributed} competed in the RC5-56 secret-key
challenge~\cite{rcs-56_url}. In 1999, the SETI@Home project~\cite{seti_url} was
released to the public to analyze radio telescope data. The Berkeley Open
Infrastructure for Network Computing (BOINC) framework~\cite{boinc_url} grew
out of SETI@Home in 2002~\cite{anderson2004boinc}, becoming the \textit{de
facto} standard framework for volunteer computing.

Volunteer computing implements a client-server architecture to achieve
high-throughput task execution. Users install a client on their own workstation
and then the client pulls tasks from the Server when CPU cycles are available.
Each client behaves as a sort of resource placeholder, one of the core features
of a \pilotjob system as seen in~\S\ref{sec:histabstr}.

HTCondor (formerly known as Condor) is a distributed computing
framework~\cite{thain2005} with a resource model similar to that of volunteer
computing. Developed around 1988, Condor enabled users to execute tasks on a
resource pool made of departmental Unix workstations. In 1996
Flocking~\cite{epema1996worldwide} implemented task scheduling over multiple
Condor resource pools and, in 2002, ``Glidein''~\cite{frey2002condorG} added
grid resources to Condor pools via resource placeholders.

Several \pilotjob systems were developed alongside Glidein to benefit from the
high-throughput and scale promised by grid resources. Around 2000,
Nimrod/G~\cite{buyya2000nimrod} extended the parameterization engine of
Nimrod~\cite{abramson1995nimrod} with resource placeholders. Four years later,
the WISDOM (wide in silico docking on malaria)~\cite{jacq2008grid} project
developed a workload manager that used resource placeholders on the EGEE
(Enabling Grids for E-Science in Europe) grid~\cite{laure2009enabling} to
compute the docking of multiple compounds, i.e. the molecules.

\jhanote{WISDOM was the name of the overall project; what was the name of the
  Pilot?}\mtnote{No stand-alone \pilotjob system was developed but a special
  purpose system composed by what we would call a workload manager and a set of
  'pilots' capable of running only two specific compound docking software. Jobs
  were run for 15 hours and multiple compounds docking were calculated for each
  job. In this way, each job acted as a placeholder and each docking calculation
  as a task executed on that placeholder. We can explain all this or try to give
  a somewhat limited summary.  I went for the latter, any better?}\jhanote{I
  inserted ``project''. OK?}

The success of grid-based \pilotjob systems and especially of Glidein
reinforced the relevance of resource placeholders to enable scientific
computation but their implementation also highlighted two main challenges:
user/system layer isolation, and application development model. For example,
Glidein allowed for the user to manage resource placeholders directly but
machine administrators had to manage the software required to create the
resource pools. Application-wise, Glidein enabled integration with application
frameworks but did not programmatically support the development of applications
by means of dedicated APIs and libraries.

Concomitant and correlated with the development of LHC~\cite{lhc_url} there was
a ``Cambrian Explosion'' of \pilotjob systems. Approximately between 2001 and
2006, five major \pilot systems were developed: DIstributed ANalysis
Environment (DIANE)~\cite{diane_date_url,diane_url}, ALIce ENvironmen
(AliEn)~\cite{saiz2003alien,alien_url}, Distributed Infrastructure with Remote
Agent Control (DIRAC)~\cite{tsaregorodtsev2003dirac,dirac_url}, Production and
Distributed Analysis (\panda)~\cite{panda_url}, and Glidein Workload Management
System (GlideinWMS)~\cite{glideinwms_date_url,glideinwms_url}. These \pilotjob
systems were developed to serve user communities and experiments at the LHC:
DIRAC is being developed and maintained by the LHCb experiment~\cite{lhcb_url};
AliEn by ALICE~\cite{collaboration2008alice}; \panda by ATLAS; and
Glide\-in\-WMS by the US national group~\cite{uscms_url} of the CMS
experiment~\cite{chatrchyan2012observation}.

The LHC \pilotjob systems have been designed to be functionally very similar,
work on almost the same underlying infrastructure, and serve applications with
very similar characteristics. Around 2011, these similarities enabled
Co-Pilot~\cite{buncicco2011co,harutyunyan2012cernvm} to support the execution
of resource placeholders on cloud and volunteer
computing~\cite{cern_challenge_url} resource pools for all the LHC experiments.

\pilotjob systems development continued to support research, resources,
middleware, and frameworks independent from the LHC experiments. ToPoS (Token
Pool Server)~\cite{topos_url} was developed around 2009 by SARA (Stichting
Academisch Rekencentrum Amsterdam)~\cite{sara_url}. ToPoS mapped tasks to
tokens and distributed tokens to resource placeholders. A REST interface was
used to store task definitions avoiding the complexities of the middleware of
high-performance machines~\cite{topos_manual_url}.

Developed around 2011, BigJob ~\cite{saga_bigjob_condor_cloud} (now
re-implemented as RADICAL-Pilot~\cite{review_radicalpilot_2015}) supported
task-level parallelism on HPC machines. BigJob extended pilots to also hold
data resources exploring the notion of
``pilot-data''~\cite{pilot-data-jpdc-2014} and uses an interoperability library
called ``SAGA'' (Simple API for Grid Applications) to work on a variety of
computing infrastructures~\cite{saga-x,goodale2006,saga_bigjob_condor_cloud}.
BigJob also offered application-level programmability of distributed
applications and their execution.

GWPilot~\cite{rubio2012performance} built upon the GridWay
meta-scheduler~\cite{huedo2007modular} to implement efficient and reliable
scheduling algorithms. Developed around 2012, GWPilot was specifically aimed at
grid resources and enabled customization of scheduling at the application
level, independent from the resource placeholder implementation.

\pilotjob systems have also been used to support science workflows. For
example, Corral~\cite{rynge2011experiences} was developed as a frontend to
Glidein and to optimize the placement of glideins (i.e., resource placeholders)
for the Pegasus workflow system~\cite{deelman2015}. Corral was later extended
to also serve as one of the frontends of GlideinWMS.
BOSCO~\cite{weitzel2012campus}, also a workflow management system, was
developed to offer a unified job submission interface to diverse middleware,
including the Glidein and GlideinWMS \pilotjob systems. The
Coaster~\cite{hategan2011coasters,coasters_url} and
Falkon~\cite{raicu2007falkon} \pilotjob systems were both tailored to support
the execution of workflows specified in the Swift
language~\cite{wilde2011swift}.


\jhanote{Should we use the past tense in Section 2, as it is a historical
  retrospective?}\mtnote{This is a tricky one because some of these systems are
  still being developed and maintained. I have changed present to past and to
  present perfect. Please check as I do not trust my grammar skills at all!}

\newcommand{\vocab}[1]{\textbf{#1}\xspace}
\newcommand{\prop}[1]{\textit{#1}\xspace}
\newcommand{\impterm}[1]{\texttt{#1}\xspace}

\section{The Pilot Abstraction}
\label{sec:understanding}

The overview presented in~\S\ref{sec:history} shows a degree of heterogeneity
among \pilotjob systems. These systems are implemented to support specific use
cases by executing certain types of workload on machines with particular
middleware. Implementation details hide the commonalities and differences among
\pilotjob systems. Consequently, in this section we describe the components,
functionalities, and architecture pattern shared by  \pilotjob systems.
Together, these elements comprise what we call the ``pilot abstraction''.

\pilotjob systems are developed by independent projects and described with
inconsistent terminologies. Often, the same term refers to multiple concepts or
the same concept is named in different ways. We address this source of
confusion by defining a terminology that can be used consistently across
\pilotjob systems, including the workloads they execute and the resources they
use.

\subsection{Logical Components and Functionalities}
\label{sec:compsandfuncs}

\begin{figure}[t]
    \centering
        \includegraphics[width=.48\textwidth]{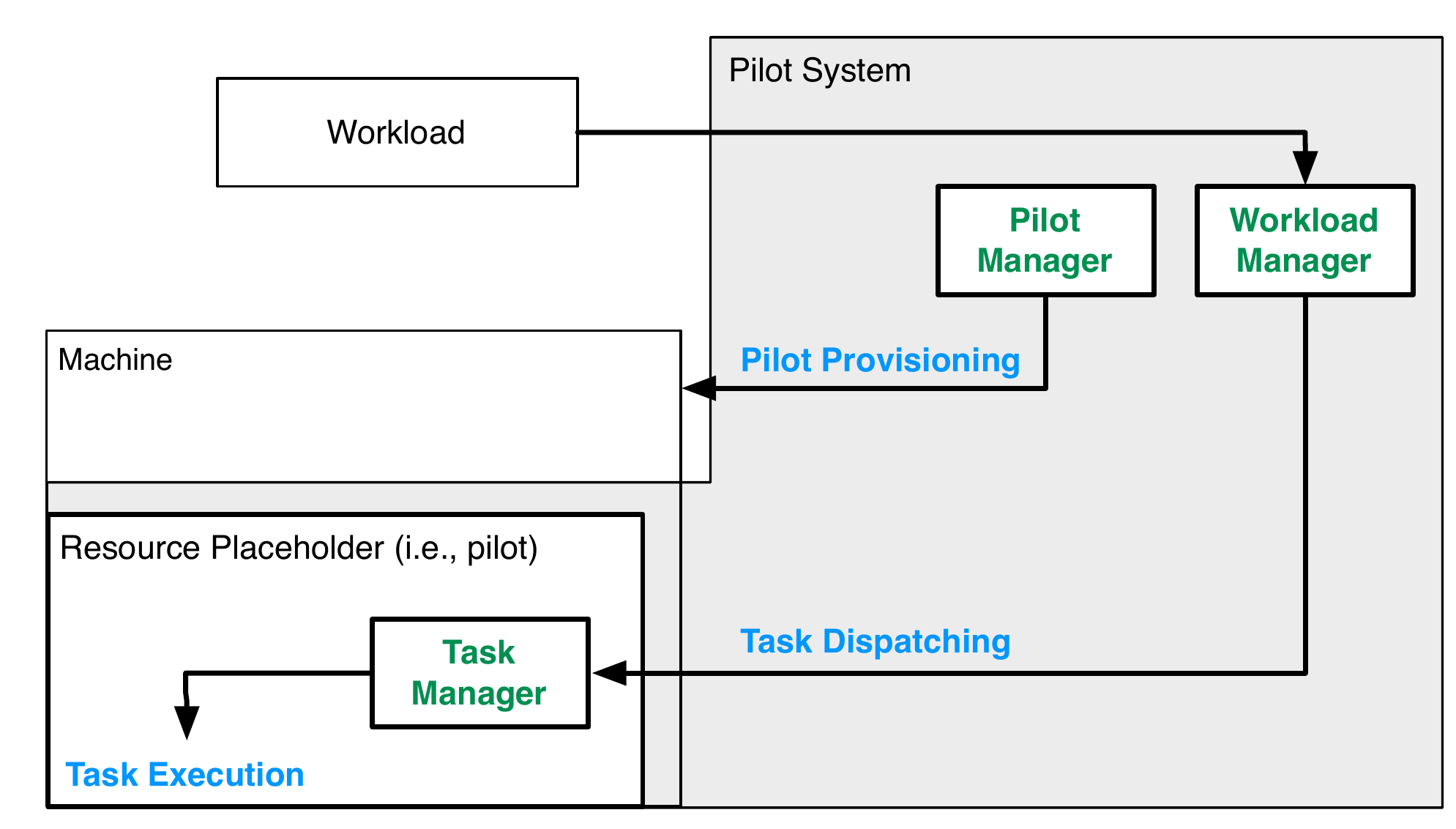}
    \caption{Diagrammatic representation of the logical components and
    functionalities of \pilot systems. The logical components are highlighted
    in green, and the functionalities in blue.}
    \label{fig:comp_func}
\end{figure}

\pilotjob systems employ three separate but coordinated logical components: a
\vocab{Pilot Manager}, a \vocab{Workload Manager}, and a \vocab{Task Manager}
(Figure~\ref{fig:comp_func}). The Pilot Manager handles the provisioning of one
or more resource placeholders (i.e., pilots) on single or multiple machines.
The Workload Manager handles the dispatching of one or more workloads on the
available resource placeholders. The Task Manager handles the execution of the
tasks of each workload on the resource placeholders.

The implementation of these three logical components vary across \pilotjob
systems (see~\S\ref{sec:analysis}). For example, two or more logical components
may be implemented by a single software element or additional functionalities
may be integrated into the three management components.

The three logical components support the common functionalities of \pilotjob
systems: \vocab{Pilot Provisioning}, \vocab{Task Dispatching}, and \vocab{Task
Execution} (Figure~\ref{fig:comp_func}). \pilotjob systems have to provision
resource placeholders on the target machines, dispatch tasks on the available
placeholders, and use these placeholders to execute the tasks of the given
workload. More functionalities may be needed to implement a production-grade
\pilotjob system as, for example, authentication, authorization, accounting,
data management, fault-tolerance, or load-balancing. However, these
functionalities depend on the type of use cases, workloads, or resources and,
as such, are not necessary to every \pilotjob system.

As seen in~\S\ref{sec:history}, resource placeholders enable tasks to utilize
resources without directly depending on the capabilities exposed by the target
machines. Resource placeholders are scheduled onto target machines by means of
dedicated capabilities, but once scheduled and then executed, these
placeholders make their resources directly available for the execution of the
tasks of a workload.

The provisioning of resource placeholders depends on the capabilities exposed
by the middleware of the targeted machine and on the implementation of each
\pilotjob system. Provisioning a placeholder on middleware with queues, batch
systems and schedulers, typically involves the placeholder being submitted as a
job. For such middleware, a job is a type of logical container that includes
configuration and execution parameters alongside information on the application
to be executed on the machine's compute resources. Conversely, for machines
without a job-based middleware, a resource placeholder might be executed by
means of other types of logical container as, for example, a virtual
machine~\cite{bernstein2014,felter2014}.


Once placeholders control a portion of a machine resources, tasks need to be
dispatched to those placeholders for execution. Task dispatching is controlled
by the \pilotjob system, not by the targeted machine's middleware. This is a
defining characteristic of \pilotjob systems because it decouples the execution
of a workload from the need to submit its tasks via the machine's scheduler.
Execution patterns involving task and/or data dependences can thus be
implemented independent of the constraints of the target machine's middleware.
Ultimately, this is how \pilotjob systems can improve workload execution
compared to direct submission.

The three logical components of a \pilotjob system -- Workload Manager, Pilot
Manager, and Task Manager -- need to communicate and coordinate in order to
execute the given workload. Any suitable communication and coordination
pattern~\cite{agha1997abstracting,omicini2013coordination} can be used and this
pattern may be implemented by any suitable technology. In a distributed
context, different network architectures and protocols may also be used to
achieve effective communication and coordination.


As seen in~\S\ref{sec:history}, \MW is a common coordination pattern among
\pilotjob systems. Workload and task Managers are implemented as separated
modules, one acting  as master and the other as worker. The master dispatches
tasks while the workers execute them independent of each other. Alternative
coordination patterns can be used where, for example, Workload and Task
Managers are implemented as a single module sharing dispatching and execution
responsibilities.


Data management can play an important role within a \pilotjob system as most of
workloads require reading input and writing output data. The mechanisms used to
make input data available and to store and share output data depend on use
cases, workloads, and resources. Accordingly, data capabilities other than
reading and writing files like, for example, data replication, (concurrent)
data transfers, non file-based data abstractions, or data placeholders should
be considered special-purpose capabilities, not characteristic of every
\pilotjob system.

\subsection{Terms and Definitions}
\label{sec:termsdefs}

In this subsection, we define a minimal set of terms related to the logical
components and capabilities of \pilotjob systems. The terms ``pilot'' and
``job'' need to be understood in the context of machines and middleware used by
\pilotjob systems. These machines offer compute, storage, and network resources
and pilots allow for the utilization of those resources to execute the tasks of
one or more workloads.

\begin{description}

\item[Task.] A set of operations to be performed on a computing platform,
alongside a description of the properties and dependences of those operations,
and indications on how they should be executed and satisfied. Implementations
of a task may include wrappers, scripts, or applications.


\item[Workload.] A set of tasks, possibly related by a set of arbitrarily
complex relations. For example, relations may involve tasks, data, or runtime
communication requirements.

\end{description}

The tasks of a workload can be homogeneous, heterogeneous, or one-of-a-kind. An
established taxonomy for workload description is not available. We propose a
taxonomy based upon the orthogonal properties of coupling, dependency, and
similarity of tasks.

Workloads comprised of tasks that are independent and indistinguishable from
each other are commonly referred to as a Bag-of-Tasks
(BoT)~\cite{da2003trading,cirne2003running}. Ensembles are workloads where the
collective outcome of the tasks is relevant (e.g., computing an average
property)~\cite{raicu2008many}. The tasks that comprise the workload in turn
can have varying degrees and types of coupling; coupled tasks might have global
(synchronous) or local (asynchronous) exchanges, and regular or irregular
communication. We categorize such workloads as coupled ensembles independent of
the specific details of the coupling between the tasks. A workflow represents a
workload with arbitrarily complex relationships among the tasks, ranging from
dependencies (e.g., sequential or data) to coupling between the tasks (e.g.,
frequency or volume of exchange)~\cite{taylor2007workflows}.

\begin{description}

\item[Resource.] A description of a finite, typed, and physical entity utilized
  when executing the tasks of a workload. Compute cores, data storage space, or
  network bandwidth between a source and a destination are examples of
  resources commonly utilized when executing workloads.

\item[Distributed Computing Resource (DCR).] A system\newline characterized by:
  a set of possibly heterogeneous resources, a middleware, and an
  administrative domain. A cluster is an example of a DCR: it offers sets of
  compute, data, and network resources; it deploys a middleware as, for
  example, the Torque batch system, the Globus grid middleware, or the
  OpenStack cloud platform; and enforces policies of an administrative domain
  like XSEDE, OSG, CERN, NERSC, or a University. So called supercomputers or
  workstations can be other examples of DCR, where the term ``distributed''
  refers to (correlated) sets of independent types of resources.

\item[Distributed Computing Infrastructure (DCI).] A set of DCRs federated with
  a common administrative, project, or policy domain, also shared at the
  software level. The federation and thus the resulting DCI can be dynamic, for
  example, a DCR that is part of XSEDE can be federated with a DCR that is part
  of OSG without having to integrate entirely the two administrative domains.




\end{description}


Our definitions of resource and DCR might seem restrictive or inconsistent with
how the term ``resource'' is sometimes used in the field of distributed
computing.  This is because the terms ``DCR'' and ``resource'' as defined here
refer to the types of machine and to the types of computing resource they
expose to the user. In its common use, the term ``resource'' conflates these
two elements because it is used to indicate specific machines like, for
example, \texttt{Stampede}, but also a specific computing resource as, for
example, compute cores.

The term ``DCR'' also offers a more precise definition of the generic term
``machine''. DCR indicates a type of machine in terms of its resources,
middleware, and administrative domain. These three elements are required to
characterize \pilotjob systems as they determine the type of resources that can
be held by a pilot, the pilot properties and capabilities, and the
administrative constraints on its instantiation.

The use of the term ``distributed'' in DCR makes explicit that the aggregation
of diverse types of resources may happen at a physical or logical level, and at
an arbitrary scale. This is relevant because the set of resources of a DCR can
belong to a physical or virtual machine as much as to a set of these
entities~\cite{sterling2002beowulf,awshpc_url,nomad_url}, either co-located on
a single site or distributed across multiple sites. Both a physical cluster of
compute nodes and a logical cluster of virtual machines are DCRs as they have a
set of resources, a middleware, and an administrative domain. 

The term ``DCI'', commonly used to indicate a distributed computing
infrastructure, is consistent with both ``resource'' and ``DCR'' as defined
here. Diverse types of resource are collected into one or more DCR, and
aggregates of DCRs that share some common administrative aspects or policy form
a DCI.


As seen in~\S\ref{sec:history}, most of the DCRs used by \pilotjob systems
utilize ``queues'', ``batch systems'', and ``schedulers''. In these DCRs, jobs
are scheduled and then executed by a batch system.

\begin{description}

\item[Job.] A type of container used to acquire resources on a DCR.

\end{description}

When considering \pilotjob systems, jobs and tasks are functionally analogous
but qualitatively different. Functionally, both jobs and tasks are containers,
i.e. metadata wrappers around one or more executables often called
``application'' or ``script''.  Qualitatively, tasks are the functional units
of a workload, while jobs are what is scheduled on a DCR. Given their
functional equivalence, the two terms can be adopted interchangeably when
considered outside the context of \pilotjob systems.

As described in~\S\ref{sec:compsandfuncs}, a resource placeholder needs to be
submitted to a DCR in order to acquire resources for the \pilotjob. The
placeholder needs to be wrapped in a container, e.g., a job, and that container
needs to be supported by the middleware of the target DCR. For this reason, the
capabilities exposed by the middleware of the target DCR determine the
submission process of resource placeholders and its specifics.

\begin{description}

\item[Pilot.] A container (e.g., a ``job'') that functions as a resource
placeholder on a given infrastructure and is capable of executing tasks of a
workload on that resource.

\end{description}

A pilot is a resource placeholder that holds portion of a DCR's resources. A
\pilotjob system is software capable of creating pilots so as to gain exclusive
control over a set of resources on one or more DCRs, and then to execute the
tasks of one or more workloads on those pilots.


The term ``pilot'' as defined here is named differently across \pilotjob
systems. In addition to the term ``placeholder'', pilots have also been named
``job agent'', ``job proxy'', ``coaster'', and ``glidein''~\cite{moscicki2011,pinchak2002,hategan2011coasters,sfiligoi2008glideinwms}.
These terms are used as synonyms, often without distinguishing between the type
of container and the type of executable that compose a pilot.

Until now, the term ``\pilotjob system'' has been used to indicate those
systems capable of executing workloads on pilots. For the remainder of this
paper, the term ``\pilot system'' will be used instead, as the term ``job'' in
``\pilotjob'' identifies just the way in which a pilot is provisioned on a DCR
exposing specific middleware. The use of the term ``\pilotjob system'' should
be regarded as a historical artifact, indicating the use of middleware in which
the term ``job'' was, and still is, meaningful.

We have now defined resources, DCRs, and pilots. We have established that a
pilot is a placeholder for a set of DCR's resources. When combined, the
resources of multiple pilots form a resource overlay. The pilots of a resource
overlay can potentially be distributed over distinct DCRs.

\begin{description}

\item[Resource Overlay.] The aggregated set of resources of multiple pilots
possibly instantiated on multiple DCRs.

\end{description}


As seen in~\S\ref{sec:histabstr}, three more terms associated with \pilot
systems need to be explicitly defined: ``early binding'', ``late binding'', and
``multi-level scheduling''.

The terms ``binding'' and ``scheduling'' are often used interchangeably but
here we use ``binding'' to indicate the association of a task to a pilot and
``scheduling'' to indicate the enactment of that association. Binding and
scheduling may happen at distinct points in time and this helps to expose the
difference between early and late binding, and multi-level scheduling.

The type of binding of tasks to pilots depends on the state of the pilot. A
pilot is inactive until it is executed on a DCR, is active thereafter, until it
completes or fails. Early binding indicates the binding of a task to an
inactive pilot; late binding the binding of a task to an active one.

Early binding is useful because by knowing in advance the properties of the
tasks that are bound to a pilot, specific deployment decisions can be made for
that pilot. For example, a pilot can be scheduled onto a specific DCR, because
of the capabilities of the DCR or because the data required by the tasks are
already available on that DCR. Late binding is instead critical to assure high
throughput by enabling sustained task execution without additional queuing time
or pilot instantiation time.

Once tasks have been bound to pilots, \pilot systems are said to implement
multi-level
scheduling~\cite{rubio2015gwpilot,de2014panda,balderrama2012scalable} because
they include scheduling onto the DCR as well as scheduling onto the pilots.
Unfortunately, the term ``level'' in multi-level is left unspecified making
unclear what is scheduled and when. Assuming the term ``entity'' indicates what
is scheduled, and the term ``stage'' the point in time at which the scheduling
happens, ``multi-entity'' and ``multi-stage'' are better terms to describe the
scheduling properties of \pilot systems. ``Multi-entity'' indicates that (at
least) two entities are scheduled and ``multi-stage'' that such scheduling
happens at separate moments in time. \pilot systems schedule pilots on DCR and
tasks on pilots at different point in time.

\begin{description}

\item[Early binding.] Binding one or more tasks to an inactive pilot.

\item[Late binding.] Binding one or more tasks to an active pilot.

\item[Multi-entity and Multi-stage scheduling.] Scheduling pilots onto
resources, and scheduling tasks onto (active or inactive) pilots.

\end{description}

\begin{figure}[t]
    \centering
        \includegraphics[width=.48\textwidth]{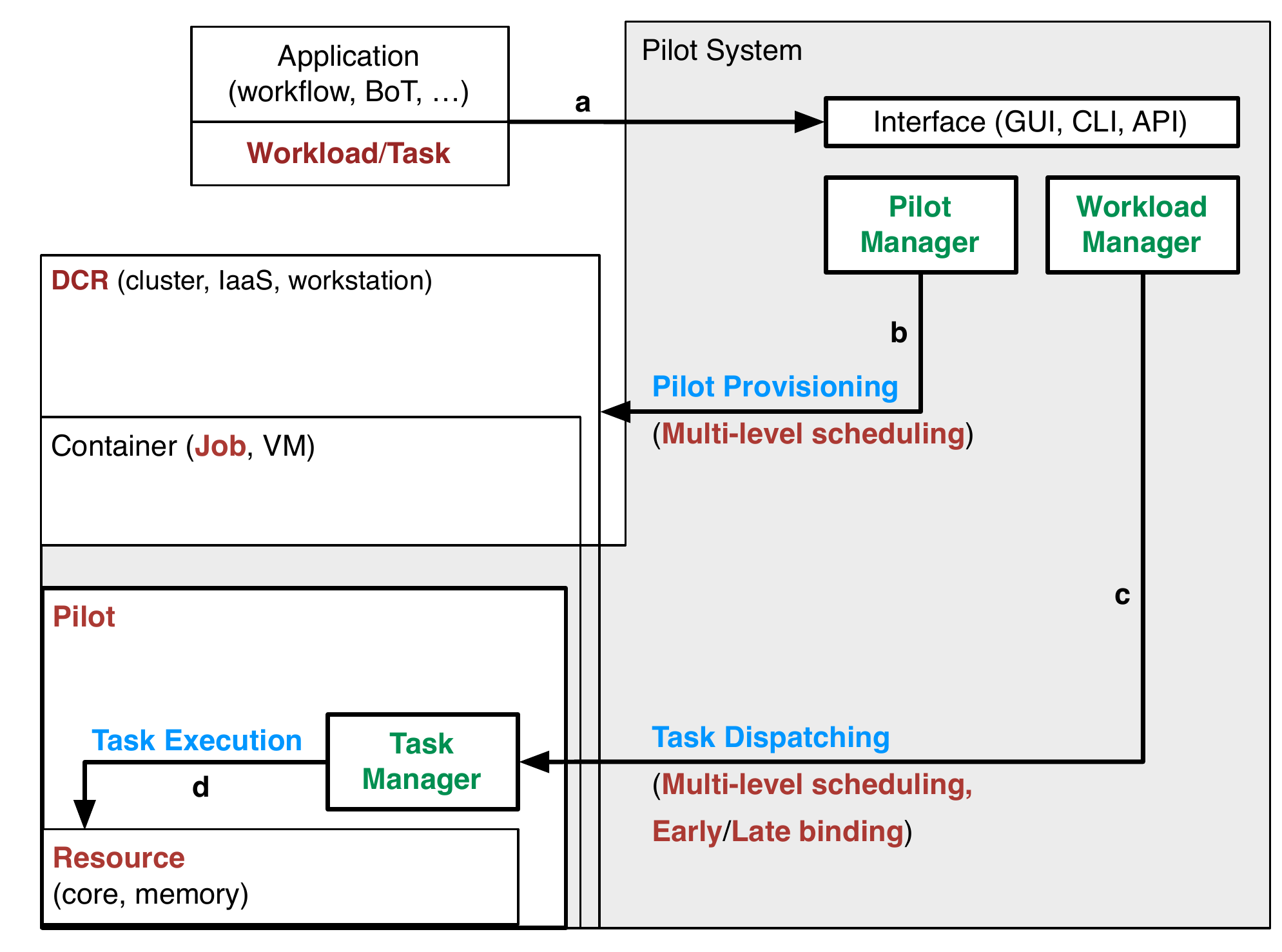}
    \caption{Diagrammatic representation of the logical components,
    functionalities, and core terminology of a \pilot system. The core
    terminology is highlighted in red, the logical components of a \pilot
    system in green, and the functionalities in blue. The components of a
    \pilot system are represented by boxes with a thicker border.}
    \label{fig:core_terminology}
\end{figure}

Figure~\ref{fig:core_terminology} offers a diagrammatic overview of the logical
components of \pilot systems (green) alongside their functionalities (blue) and
the defined terminology (red). The figure is composed of three main blocks: the
one on the top-left corner represents the workload originator. The one starting
at the top-right and shaded in gray represents the \pilot system, while the
four boxes one inside the other on the left side of the figure represent a DCR.
Of the four boxes, the outmost denotes the DCR boundaries, e.g., a cluster. The
second box the container used to schedule a pilot on the DCR, e.g., a job or a
virtual machine. The third box represents the pilot once it has been
instantiated on the DCR, and the fourth box represents the resources held by
the pilot. The boxes representing the components of a \pilot system have been
highlighted with a thicker border.

Figure~\ref{fig:core_terminology} shows the separation between the DCR and the
\pilot system, and how the resources on which tasks are executed are contained
in the DCR within different logical and physical components. Appreciating the
characteristics and functionalities of a \pilot system depends upon
understanding the levels at which each of its component exposes capabilities.
An application submits one or more workloads composed of tasks to the \pilot
system via an interface (tag~a). The Pilot Manager is responsible for pilot
provisioning (tag~b), the Workload Manager to dispatch tasks to the Task
Manager (tag~c), the Task Manager to execute those tasks once the pilot has
become available (tag~d).

Note how in Figure~\ref{fig:core_terminology} scheduling happens at the DCR
(tag~b), for example, by means of a cluster scheduler, and then at the pilot
(tag~c). This illustrates what here has been called ``multi-entity'' and
``multi-stage'' scheduling, replacing the more common but less precise term
multi-level scheduling. The separation between scheduling at the pilot and
scheduling at the Workload Manager highlights the four entities involved in the
two-stage scheduling: jobs on DCR middleware, and tasks on pilots. This helps
to appreciate the critical distinction between the container of a pilot and the
pilot itself. A container is used by the Pilot Manager to provision the pilot.
Once the pilot has been provisioned, it is the pilot and not the container that
is responsible of both holding a set of resources and offering the
functionalities of the Task Manager.

Figure~\ref{fig:core_terminology} should not be confused with an architectural
diagram. No indications are given about the interfaces that should be used, how
the logical component should be mapped into software modules, or what type of
communication and coordination protocols should be adopted among such
components. This is why no distinction is made diagrammatically between, for
example, early and late binding.

Figure~\ref{fig:core_terminology} is instead an architectural
pattern~\cite{buschmann2007pattern} for systems that execute workloads on
multiple DCRs via pilot-based multi-entity, many-stage scheduling of tasks.
This pattern can be realized into an architectural description and then
implemented into a specific \pilot system. Several architectural models,
frameworks, languages, supporting platforms, and standards are available to
produce architectural
descriptions~\cite{kruchten2006past,kazman2014software}. Common examples are 4+1
architectural view~\cite{kruchten19954}, Open Distributed Processing
(ODP)~\cite{raymond1995reference}, Zachman~\cite{zachman2006zachman}, The Open
Group Architecture Framework (TOGAF)~\cite{josey2011togaf}, and the
Attribute-Driven Design (ADD)~\cite{wojcik2006attribute}.

\section{Pilot Systems}
\label{sec:analysis}

In this section we examine multiple implementations of \pilot systems.
Initially, we derive core and auxiliary properties of \pilot system
implementations from the components and functionalities described
in~\S\ref{sec:compsandfuncs}. Subsequently, we describe a selection of \pilot
system implementations showing how the architecture of each system maps to the
architectural pattern presented in~\S\ref{sec:termsdefs}. Finally, we offer
insight about the commonalities and differences among the described \pilot
system implementations discussing also their most relevant auxiliary
properties.

%
\subsection{Core properties}
\label{sec:coreprops}

Core properties are specific to \pilot systems and necessary for their
implementation. These properties characterize \pilot systems because they
relate to pilots and how they are used to execute tasks. Without core
properties Pilot Managers, Workload Managers, and Task Managers would not be
capable to provide pilots, and to dispatch and execute tasks. We list the core
properties of \pilot systems in Table~\ref{table:core_properties}.

The first three core properties -- Pilot Scheduling, Pilot Bootstrapping, and
Pilot Resources -- relate to the procedures used to provision pilots and to the
resources they hold. Pilots can be deployed by a Pilot Manager using a suitable
wrapper that can be scheduled on the targeted DCR middleware. Pilots become
available only with a correct bootstrapping procedure, and they can be used for
task execution only if they acquire at least one type of resource, e.g.,
compute cores or data storage.


\begin{table*}
\centering
\begin{tabular}{p{3.3cm}p{7.6cm}p{2.6cm}p{2.5cm}}

\toprule

\textbf{Property} &
\textbf{Description} &
\textbf{Component} &
\textbf{Functionality} \\

\midrule

Pilot Scheduling &
Modalities for pilot scheduling on DCRs &
\multirow{2}{*}{Pilot Manager} &
\multirow{2}{*}{Pilot Provisioning} \\

Pilot Bootstrapping &
Modalities for pilot bootstrapping on DCRs &
&
\\

Pilot Resources &
Types and characteristics of pilot resources &
&
\\


\midrule

Workload Binding &
Modalities and policies for binding tasks to pilots &
\multirow{2}{*}{Workload Manager} &
\multirow{2}{*}{Task Dispatching} \\

Workload Scheduling &
Modalities and policies for scheduling tasks to pilots &
&
\\

\midrule

Workload Environment &
Type and features of the task execution environment &
Task Manager &
Task Execution \\

\bottomrule

\end{tabular}
\caption{\textbf{Mapping of the core properties  of \pilot system
  implementations onto the components and functionalities described in
  \S\ref{sec:compsandfuncs}. Core properties are specific to \pilot systems and
  necessary for their implementation.} }
\label{table:core_properties}
\end{table*}

The Workload Binding and Workload Scheduling core properties relate to how
\pilot systems bind tasks to pilots, and then how these tasks are scheduled
once pilots become available. A Workload Manager can early or late bind tasks
to pilots depending on the DCR's resources and workload's requirements.
Scheduling decisions may depend on the number and capabilities of the available
pilots or on the status of workload execution. Workload Binding and Workload
Scheduling enable \pilot systems to control the coupling between tasks
requirements and pilot capabilities.

The Workload Environment core property relates to the features and configuration
of the environment provided by the pilot in which tasks are executed on the DCR.
A Task Manager requires information about the environment to successfully manage
the execution of tasks. For example, the Task Manager may have to make available
supporting software or choose suitable parameters for the task executable.  The
following describes each core property. Note that these properties refer to
\pilot systems and not to individual pilots instantiated on a DCR.


\begin{itemize}

\item \textbf{Pilot Scheduling}. Modalities for scheduling pilots on DCRs.
  Pilot scheduling may be: fully automated (i.e., implicit) or directly
  controlled by applications or users (i.e., explicit); performed on a single
  DCR (i.e., local) or coordinated across multiple DCRs (i.e., global);
  tailored to the execution of the workload (i.e., adaptive) or predefined on
  the basis of policies and heuristics (i.e, static).

\item \textbf{Pilot Bootstrapping}. Modalities for pilot bootstrapping on DCRs.
  Pilots can be bootstrapped from code downloaded at every instantiation or
  from code that is bundled by the DCR. The design of pilot bootstrapping
  depends on the DCR environment and on whether single or multiple types of
  DCRs are targeted. For example, a design based on connectors can be used with
  multiple DCRs to get information about container type (e.g., job, virtual
  machine), scheduler type (e.g., PBS, HTCondor, Globus), amount of cores,
  walltime, or available filesystems.

\item \textbf{Pilot Resources}. Types and characteristics of the resources
  exposed by a \pilot system. Resource types are, for example, compute, data,
  or networking while some of the their typical characteristics are: size
  (e.g., number of cores or storage capacity), lifespan, intercommunication
  (e.g., low-latency or inter-domain), computing platforms (e.g., x86 or GPU),
  file systems (e.g., local or distributed). The resource held by a pilot
  varies depending on the system architecture of the DCR in which the pilot is
  instantiated.  For example, a pilot may hold multiple compute nodes, single
  nodes, or portion of the cores of each node. The same applies to file systems
  and their partitions or to physical and software-defined networks.

\item \textbf{Workload Binding}. Time of workload assignment to pilots.
  Executing a workload requires its tasks to be bound to one or more pilots
  before or after they are instantiated on a DCR. As seen
  in~\S\ref{sec:understanding}, \pilot systems may allow for two modalities of
  binding between tasks and pilots: early binding and late binding. \pilot
  system implementations differ in whether and how they support these two types
  of binding. 


\item \textbf{Workload Scheduling}. Enactment of a binding. \pilot systems can
  support (prioritized) application-level or multi-stage scheduling decisions.
  Coupled tasks may have to be scheduled on a single pilot, loosely coupled or
  uncoupled tasks to multiple pilots; tasks may be scheduled to a pilot and
  then to a specific pool of resources on a single compute node; or task
  scheduling may be prioritized depending on task size and duration.

\item \textbf{Workload Environment}. Type, dependences, and characteristics of
  the environment in which workload's tasks are executed. Once scheduled to a
  pilot, a task needs an environment that satisfies its execution requirements.
  The execution environment depends on the type of task (e.g., single or
  multi-threaded, MPI), task code dependences (e.g., compilers, libraries,
  interpreters, or modules), and task communication, coordination and data
  requirements (e.g., interprocess, inter-node communication, data staging,
  sharing, and replication).

\end{itemize}

\jhanote{I notice that in the paper we don't use a ``,'' before conjunctions.
  The only time when a ``,'' is not used with a conjunction is when it is a
  subordinate conjunction: https://owl.english.purdue.edu/engagement/2/1/37/.
  Are you OK, if in my second pass, I add a ``,'' before conjunction of
  independent clauses.}\mtnote{Absolutely, thank you for the link!}

%
\subsection{Auxiliary properties}
\label{sec:auxprops}

Auxiliary properties are not specific to \pilot systems and may be optional for
their implementation. \pilot systems share auxiliary properties with other
types of system and \pilot system implementations may have different subsets of
these properties. For example, authentication and authorization are properties
shared by many systems and \pilot systems may have to implement them only for
some DCRs. Analogously, communication and coordination is not a core property
of \pilot systems because, at some level, all software systems require
communication and coordination.

We list a representative subset of auxiliary properties for \pilot systems in
Table~\ref{table:aux_properties}. The following describes these auxiliary
properties and, also in this case, these properties refer to
\pilot systems and not to individual pilots instantiated on a DCR.

\begin{table*}
\centering
\begin{tabular}{p{5cm}p{11cm}}

\toprule

\textbf{Property}      &
\textbf{Description}\\

\midrule

Architecture &
Structures and components of the \pilot system \\

Coordination and Communication &
Interaction protocols and patterns among the components of the system \\

Interface &
Interaction mechanisms both among components and exposed to the user \\

Interoperability &
Qualitative and functional features shared among \pilots systems \\

Multitenancy &
Simultaneous use of the \pilot system components by multiple users \\

Resource Overlay &
The aggregation of resources from multiple pilots into overlays \\

Robustness &
Resilience and reliability of pilot and workload executions \\

Security &
Authentication, authorization, and accounting framework \\

Files and Data &
Mechanisms for data staging and management \\

Performance &
Measure of the scalability, throughput, latency, or memory usage \\

Development Model &
Practices and policies for code production and management \\

DCR Interaction &
Modalities and protocols for pilot system/DCR interaction coordination \\

\bottomrule

\end{tabular}
\caption{\textbf{Sample of Auxiliary Properties and their descriptions.
  Auxiliary properties are not specific to \pilot systems and may be optional
  for their implementation.}}
\label{table:aux_properties}
\end{table*}

\begin{itemize}

\item \textbf{Architecture}. \pilot systems may be implemented by means of
  different architectures, e.g., service-oriented, client-server, or
  peer-to-peer. Architectural choices may depend on multiple factors, including
  application use cases, deployment strategies, or interoperability
  requirements.

\item \textbf{Communication and Coordination}. As discussed
  in~\S\ref{sec:compsandfuncs}, \pilot system implementations are not defined
  by any specific communication and coordination protocol or pattern.
  Communication and coordination among the \pilot system components are
  determined by its design, the chosen architecture, and the deployment
  scenarios.

\item \textbf{Workload Semantics}. \pilotjob systems may support workloads with
  different compute and data requirements, and inter-task dependences. \pilot
  systems may assume that only workloads with a specific semantics are given or
  may allow the user to specify, for example, BoT, ensemble, or workflow.

\item \textbf{Interface}. \pilot systems may implement several private and
  public interfaces: among the components of the \pilot system; among the
  \pilot system, the applications, and the DCRs; or between the \pilot system
  and the users via one or more application programming interfaces.

\item \textbf{Interoperability}. \pilot system may implement at least two types
  of interoperability: among \pilot system implementations, and among DCRs with
  heterogeneous middleware. For example, two \pilot systems may execute tasks
  on each others' pilots, or a \pilot system may be able to provide pilots on
  LSF, Slurm, Torque, or OpenStack middleware.


\item \textbf{Multitenancy}. \pilot systems may offer multitenancy at both
  system and local level. When offered at system level, multiple users can
  utilize the same instance of a \pilot system; when available at local level,
  multiple users can share the same pilot. Executing multiple pilots on the
  same DCR indicates the multitenancy of the DCR, not of the \pilot system.

\item \textbf{Robustness}. Indicates the features of a \pilot system that
  contribute to its resilience and reliability. Usually, fault-tolerance,
  high-availability, and state persistence are indicators of the maturity of
  the \pilot system implementation and its use cases support.

\item \textbf{Security}. The deployment and usability of \pilot systems are
  influenced by security protocols and policies. Authentication and
  authorization can be based on diverse protocols and vary across \pilot
  systems.

\item \textbf{Data Management}. As discussed in~\S\ref{sec:compsandfuncs}, only
  basic data reading/writing functionalities are mandated by a \pilot system.
  Nonetheless, most real-life use cases require more advanced data management
  functionalities that can be implemented within the \pilot system or delegated
  to third-party tools.

\item \textbf{Performance and scalability}. \pilot systems can be optimized for
  one or more performance metrics, depending on the target use cases.  For
  example, \pilot systems vary in terms of overheads they add to the execution
  of a given workload, size and duration of the workloads a user can expect to
  be supported, and type and number of supported DCRs and DCIs.

\item \textbf{Development Model}. The model used to develop \pilot systems may
  have an impact on the life span of the \pilot system, its maintainability
  and, possibly its evolution path. This is especially relevant when
  considering whether the development is supported by an open community or by a
  single research project.

\end{itemize}

\subsection{Implementations}
\label{sec:implementations}

We analyze seven \pilot systems based on their availability, design, intended
use, and uptake. We describe systems that: (i) implement diverse design; (ii)
target specific or general-purpose use cases and DCR; and (iii) are currently
available, actively maintained, and used by scientific communities. Space
constraints prevented consideration of additional \pilot systems, as well as
necessitated limiting the analysis to the core properties of \pilot systems.

We compare \pilot systems using the architectural pattern and common
terminology defined in~\S\ref{sec:understanding}.
Table~\ref{table:implementations-components} shows how the components of the
architectural pattern are named differently across implementations.
Table~\ref{table:implementations-properties} offers instead a summary of how
the core properties are implemented for each \pilot system we
compared.\protect\footnote{Pilot systems are ordered alphabetically in the
table and in the text.}

\begin{table*}
 \centering
 \begin{tabular}{p{3cm}p{3.25cm}p{3.25cm}p{3.25cm}p{3.25cm}}

  \toprule

    \textbf{Pilot System} &
    \textbf{Pilot Manager} &
    \textbf{Workload Manager} &
    \textbf{Task Manager} &
    \textbf{Pilot} \\

  \midrule

    Coaster System &
    Coaster Service &
    Coaster Client &
    Worker &
    Job Agent \\

    DIANE &
    Submitter script &
    RunMaster &
    ApplicationWorker &
    WorkerAgent \\

    DIRAC &
    WMS (Directors) &
    WMS (Match Maker) &
    Job Wrapper &
    Job Agent \\



    GlideinWMS &
    Glidein Factory &
    Schedd &
    Startd &
    Glidein \\

    MyCluster &
    Cluster Builder Agent &
    Virtual Login Session &
    Task Manager &
    Job Proxy \\

    \panda &
    Grid Scheduler &
    PANDA Server &
    RunJob &
    Pilot \\

    RADICAL-Pilot &
    Pilot Manager &
    CU Manager &
    Agent &
    Pilot \\

 \bottomrule

 \end{tabular}
 \caption{\textbf{Mapping of the names given to the components of the pilot
 architectural pattern defined in~\S\ref{sec:termsdefs},
 Figure~\ref{fig:core_terminology} and the names given to the components of
 pilot system implementations.}
 \label{table:implementations-components}}
\end{table*}

\begin{table*}
 \centering
  \begin{tabular}{p{2.5cm}p{2.25cm}p{2cm}p{5cm}p{1.75cm}p{1.75cm}p{1.75cm}|}

  \toprule

    \textbf{Pilot\newline System} &
    \textbf{Pilot\newline Resources} &
    \textbf{Pilot\newline Deployment} &
    \textbf{Workload\newline Semantics} &
    \textbf{Workload\newline Binding} &
    \textbf{Workload\newline Execution} \\

  \midrule

    Coaster System &
    Compute &
    Implicit &
    WF (Swift~\cite{korkhov2009dynamic}) &
    Late &
    Serial, MPI \\

    DIANE &
    Compute &
    Explicit &
    WF (MOTOUR~\cite{korkhov2009dynamic}) &
    Late &
    Serial \\

    DIRAC &
    Compute &
    Implicit &
    WF (TMS) &
    Late &
    Serial, MPI \\



    GlideinWMS &
    Compute &
    Implicit &
    WF (Pegaus, DAGMan~\cite{dagman_url}) &
    Late &
    Serial, MPI \\

    MyCluster &
    Compute &
    Implicit &
    job descriptions &
    Late &
    Serial, MPI \\

    \panda &
    Compute &
    Implicit &
    BoT &
    Late &
    Serial, MPI \\

    RADICAL-Pilot &
    Compute, data &
    Explicit &
    ENS (EnsembleMD Toolkit~\cite{emdtoolkit_url}) &
    Early, Late &
    Serial, MPI \\

 \bottomrule

 \end{tabular}
 \caption{\textbf{Overview of \pilot systems and a summary of the values of
 their core properties. Based on the tooling currently available for each
 \pilot system, the types of workload supported as defined
 in~\S\ref{sec:termsdefs} are: BoT~$=$~Bag of Tasks; ENS~$=$~Ensembles;
 WF~$=$~workflows.}}
 \label{table:implementations-properties}
\end{table*}

%
\subsubsection{Coaster System}\label{sec:coaster}


The Coaster System (also referred to in literature as Coasters) was developed
by the Distributed Systems Laboratory at the University of
Chicago~\cite{dsluc_url} and it is currently maintained by the Swift
project~\cite{swift_url}. Initially developed within the CoG
project~\cite{cog_url} and maintained in a separate, standalone repository,
today the Coaster System provides pilot functionalities to Swift by means of an
abstract task interface~\cite{zhao2007swift,von2000cog}.

The Coaster System is composed of three main
components~\cite{hategan2011coasters}: a Coaster Client, a Coaster Service, and
a set of Workers. The Coaster Client implements both a Bootstrap and a
Messaging Service while the Coaster Service implements a data proxy service and
a set of job providers for diverse DCRs middleware. Workers are executed on the
DCR compute nodes to bind compute resources and execute the tasks submitted by
the users to the Coaster System.

\begin{figure}[t]
    \centering
        \includegraphics[width=.48\textwidth]{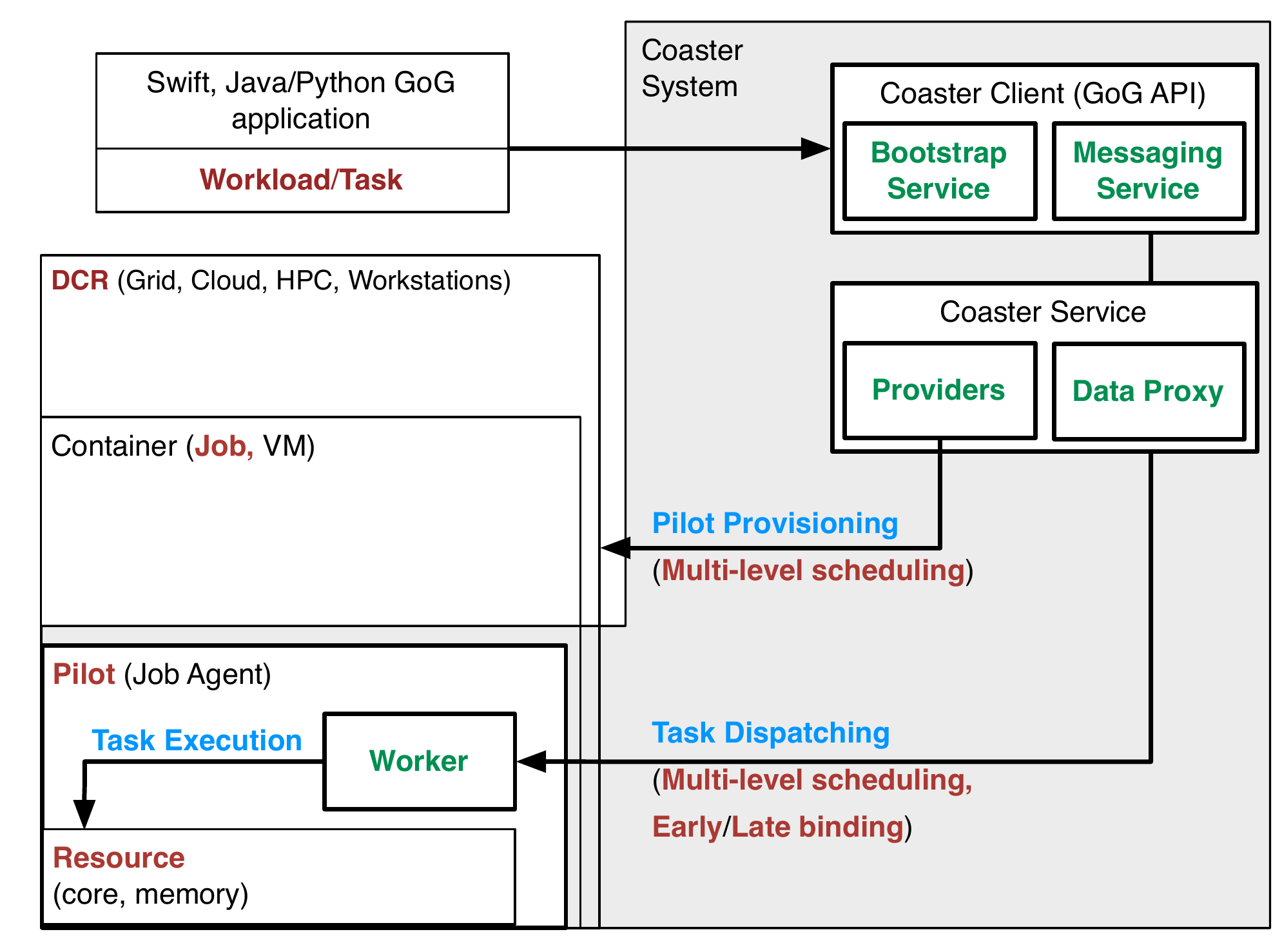}
    \caption{Diagrammatic representation of the Coaster System components,
    functionalities, and core terminology mapped on
    Figure~\ref{fig:core_terminology}.}
    \label{fig:coaster_comparison}
\end{figure}

Figure~\ref{fig:coaster_comparison} illustrates how the Coaster System
components map to the components and functionalities of a \pilot system as
described in in~\S\ref{sec:understanding}: the Coaster Client is a Workload
Manager, the Coaster Service a Pilot Manager, and each Worker a Task Manager.
The Coaster Service implements the Pilot Provisioning functionality by
submitting adequate numbers of Workers on suitable DCRs. The Coaster Client
implements Task Dispatching while the Workers implement Task Execution.

The execution model of the Coaster System can be summarized in seven
steps~\cite{coasters_url}: 1. a set of tasks is submitted by a user via the
Coaster Client API; 2. when not already active, the Bootstrap Service and the
Message Service are started within the Coaster Client; 3. when not already
active, a Coaster Service is instantiated for the DCR(s) indicated in the task
descriptions; 4. the Coaster Service gets the task descriptions and analyzes
their requirements; 5. the Coaster Service submits one or more Workers to the
target DCR taking also into account whether any other Worker is already active;
6. when a Worker becomes active it pulls a task and, if any, its data
dependences from the Coaster Client via the Coaster Service; 7. the task is
executed.


Each Worker holds compute resources in the form of compute cores. Data can be
staged from a shared file-system, directly from the client to the Worker, or
via the Coaster Service acting as a proxy. Data are not a type of resource held
by the pilots and pilots are not used to expose data to the user. Networking
capabilities are assumed to be available among the components of the Coaster
System, but a dedicated communication protocol is implemented and also used for
data staging as required.


The Coaster Service automates the deployment of pilots (i.e., Workers) by
taking into account several parameters: total number of jobs that the DCR batch
system accepts; number of cores for each DCR compute node; DCR policy for
compute nodes allocation; walltime of the pilots compared to the total walltime
of the tasks submitted by the users. These parameters are evaluated by a custom
pilot deployment algorithm that performs a walltime overallocation estimated
against user-defined parameters, and chooses the number and sizing of pilots on
the base of the target DCR capabilities.

The Coaster System serves as a \pilot backend for the Swift System and,
together, they can execute workflows composed of loosely coupled tasks with
data dependences. Natively, the Coaster Client implements a Java CoG Job
Submission Provider~\cite{von2000cog,von2005workflow,cog_url} for which Java
API are available to submit tasks and to develop distributed applications.
While tasks are assumed to be single-core by default, multi-core tasks can be
executed by configuring the Coaster System to submit Workers holding multiple
cores~\cite{swift_guide_url}. It should also be possible to execute MPI tasks
by having Workers to span multiple compute nodes of a DCR.


The Coaster Service uses providers from the Java CoG Kit Abstraction Library to
submit Workers to DCR with grid, HPC, and cloud middleware. The late binding of
tasks to pilots is implemented by Workers pulling tasks to be executed as soon
as free resources are available. It should be noted that tasks are bound to the
pilots instantiated on a specific DCR specified as part of the task
description. Experiments have been made with late binding to pilots
instantiated on arbitrary DCRs but no documentation is currently available
about the results obtained.\footnote{Based on private communication with the
Coaster System development team.}

\subsubsection{DIANE}\label{sec:diane}

DIANE (DIstributed ANalysis Environment)~\cite{moscicki2003diane} has been
developed at CERN~\cite{cern_url} to support the execution of workloads on the
DCRs federated to be part of European Grid Infrastructure (EGI)~\cite{egi_url}
and worldwide LHC Computing Grid (WLCG). DIANE has also been used in the Life
Sciences~\cite{moscicki2004biomedical,jacq2007virtual,moscicki2003} and in few
other scientific domains~\cite{bacu2011gswat,mantero2003simulation}.


DIANE is an application task coordination framework that executes distributed
applications using the \MW pattern~\cite{moscicki2003diane}.  DIANE consists of
four logical components: a TaskScheduler, an ApplicationManager, a
SubmitterScript, and a set of ApplicationWorkers~\cite{diane_api_url}. The
first two components -- TaskScheduler and the ApplicationManager -- are
implemented as a RunMaster service, while the ApplicationWorkers as a
WorkerAgent service. Submitter Scripts deploy ApplicationWorkers on DCRs.

\begin{figure}[t]
    \centering
        \includegraphics[width=.48\textwidth]{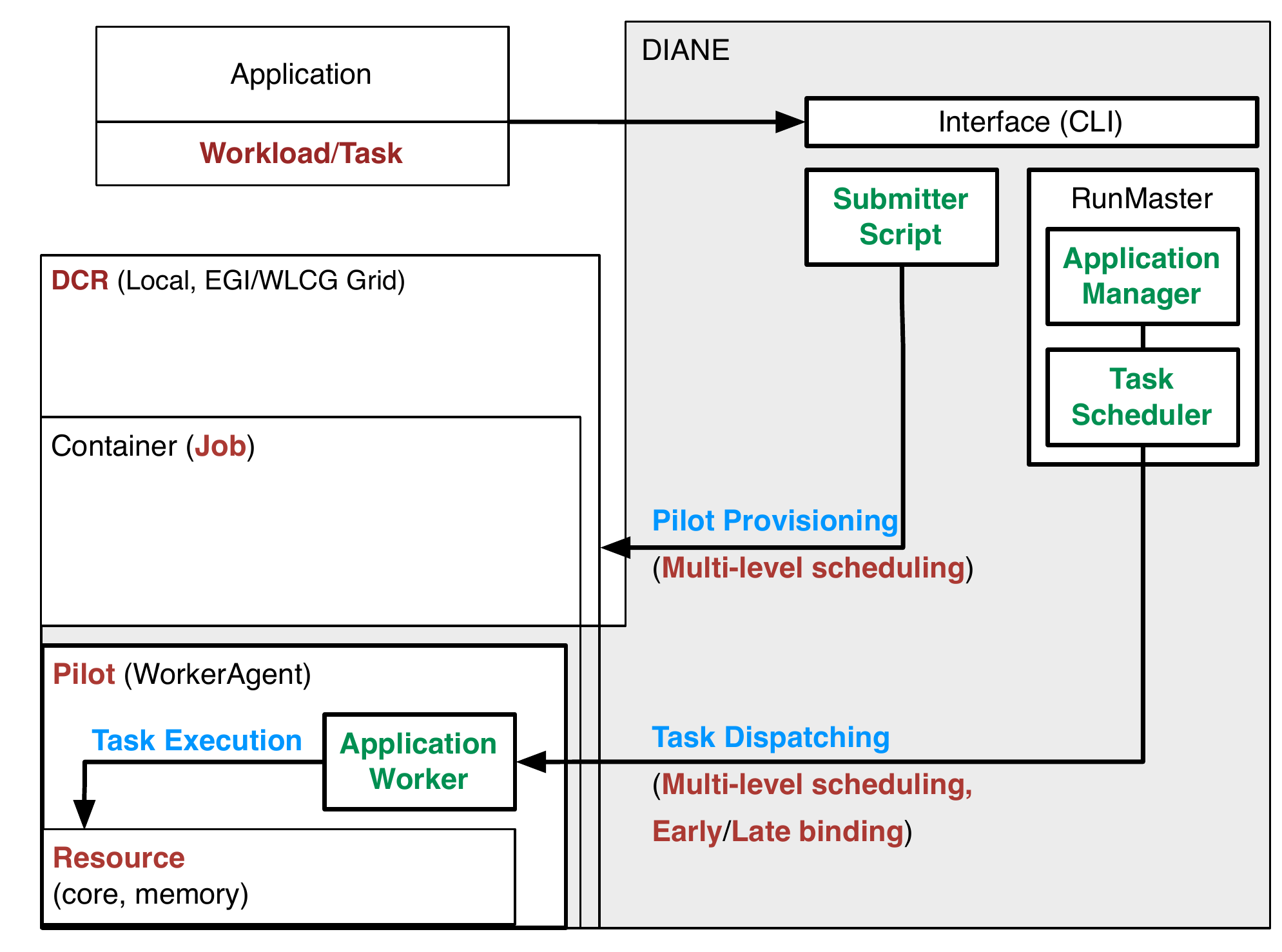}
    \caption{Diagrammatic representation of DIANE components, functionalities,
    and core terminology mapped on Figure~\ref{fig:core_terminology}.}
    \label{fig:diane_comparison}
\end{figure}

Figure~\ref{fig:diane_comparison} shows how DIANE implements the components and
functionalities of a pilot system as described in~\S\ref{sec:understanding}:
the RunMaster service is a Workload Manager, the SubmitterScript is a Pilot
Manager, and the ApplicationWorker of each WorkerAgent service is a Task
Manager. Accordingly, the \pilot provisioning functionality is implemented by
the SubmitterScript, Task Dispatching by the RunMaster, and Task Execution by
the WorkerAgent. In DIANE, Pilots are called ``WorkerAgents''.

The execution model of DIANE can be summarized in four
steps~\cite{moscicki2011understanding}: 1. the user submits one or more jobs to
DCR by means of SubmitScript(s) to bootstrap one or more WorkerAgent; 2. When
ready, the WorkAgent(s) reports back to the ApplicationManager; 3. tasks are
scheduled by the TaskScheduler on the available WorkerAgent(s); 4. after
execution, WorkerAgents send the output of the computation back to the
ApplicationManager.


The pilots used by DIANE (i.e., WorkerAgents) hold compute resources on the
target DCRs. WorkerAgents are executed by the DCR middleware as jobs with
mostly one core but possibly more. DIANE also offers a data service with a
dedicated API and CLI that allows for staging files in and out of WorkerAgents.
This service represents an abstraction of the data resources and capabilities
offered by the DCR, and it is designed to handle data only in the form of files
stored into a file system. Network resources are assumed to be available among
DIANE components.


DIANE requires a user to develop pilot deployment mechanisms tailored to
specific resources. The RunMaster service assumes the availability of pilots to
schedule the tasks of the workload.  Deployment mechanisms can range from
direct manual execution of jobs on remote resources, to deployment scripts, or
full-fledged factory systems to support the sustained provisioning of pilots
over extended periods of time.


A tool called ``GANGA''~\cite{moscicki2009ganga,ganga_url} is available to
support the development of SubmitterScripts. GANGA facilitates the submission
of pilots to diverse DCRs by means of a uniform interface and abstraction.
GANGA offers interfaces for job submission to DCRs with Globus, HTCondor,
UNICORE, or gLite middleware.

DIANE has been designed to execute workloads that can be partitioned into
ensembles of parametric tasks on multiple pilots. Each task can consist of an
executable invocation but also of a set of instructions, OpenMP threads, or MPI
processes~\cite{moscicki2011understanding}. Relations among tasks and group of
tasks can be specified before or during runtime enabling DIANE to execute
articulated workflows. Plugins have been written to manage
DAGs~\cite{grzeslo2009} and data-oriented workflows~\cite{glatard2008}.

DIANE is primarily designed for HTC and Grid environments and to execute pilots
with a single core. Nonetheless, the notion of ``capacity'' is exposed to the
user to allow for the specification of pilots with multiple cores. Although the
workload binding is controllable by the user-programmable TaskScheduler, the
general architecture is consistent with a pull model. The pull model naturally
implements the late binding paradigm where every ApplicationAgent of each
available pilot pulls a new task.

%
\subsubsection{DIRAC}\label{sec:dirac}

DIRAC (Distributed Infrastructure with Remote Agent
Control)~\cite{tsaregorodtsev2004dirac} is a software product developed by the
CERN LHCb project. DIRAC implements a Workload Management System (WMS) to
manage the processing of detector data, Monte Carlo simulations, and end-user
analyses. DIRAC primarily serves as the LHCb workload management interface to
WLCG executing workloads on DCRs deploying Grid, Cloud, and HPC middleware.

DIRAC has four main logical components: a set of TaskQueues, a set of
TaskQueueDirectors, a set of JobWrappers, and a MatchMaker. TaskQueues,
TaskQueueDirectors, and the MatchMaker are implemented within a monolithic WMS.
Each TaskQueue collects tasks submitted by users, multiple TaskQeue being
created depending on the requirements and ownership of the tasks. JobWrappers
are executed on the DCR to bind compute resources and execute tasks submitted
by the users. Each TaskQueueDirector submits JobWrappers to target DCRs. The
MatchMaker matches requests from JobWrappers to suitable tasks into TaskQueues.

\begin{figure}[t]
    \centering
        \includegraphics[width=.48\textwidth]{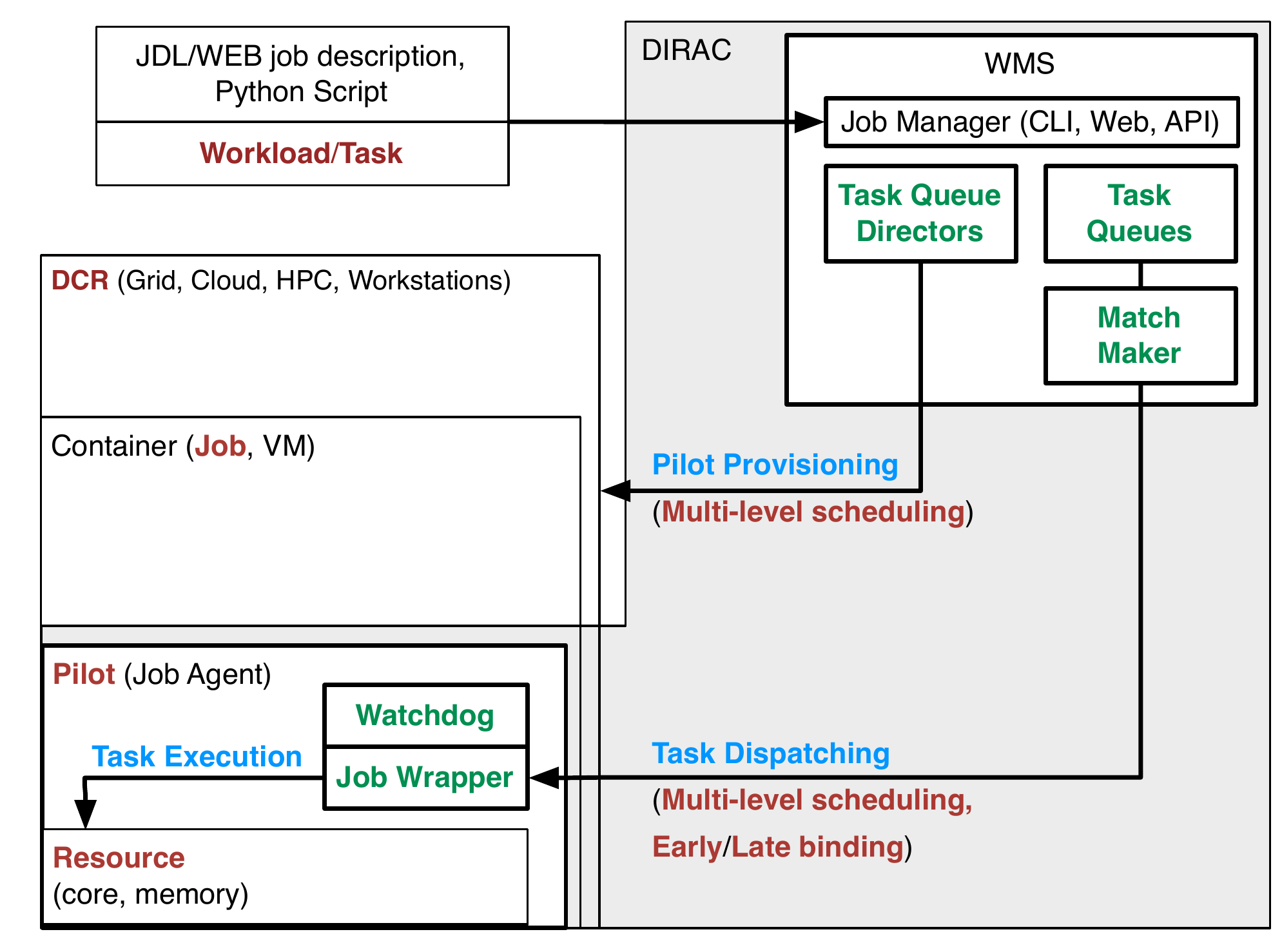}
    \caption{Diagrammatic representation of DIRAC components, functionalities,
    and core terminology mapped on Figure~\ref{fig:core_terminology}.}
    \label{fig:dirac_comparison}
\end{figure}

DIRAC was the first pilot-based WMS designed to serve a LHC main
experiment~\cite{casajus2010dirac}. Figure~\ref{fig:dirac_comparison} shows how
the DIRAC WMS implements a Workload, a Pilot, and a Task Manager as they have
been described in~\S\ref{sec:understanding}. TaskQueues and the MatchMaker
implement the Workload Manager and the related Task Dispatching functionality.
Each TaskQueueDirector implements a Pilot Manager and its Pilot Provisioning
functionality, while each JobWrapper implements a Task Manager and Pilot
Execution.

The DIRAC execution model can be summarized in five steps: 1. a user submits
one or more tasks by means of a CLI, Web portal, or API to the WMS Job Manager;
2. submitted tasks are validated and added to a new or an existing TaskQueue,
depending on the task properties; 3. one or more TaskQueues are evaluated by a
TaskQueueDirector and a suitable number of JobWrappers are submitted to
available DCRs; 4. JobWrappers, once instantiated on the DCRs, pull the
MatchMaker asking for tasks to be executed; 5. tasks are executed by the
JobWrappers under the supervision of each JobWrapper's Watchdog.


JobWrappers, the DIRAC pilots, hold compute resources in the form of single or
multiple cores, spanning portions, whole, or multiple compute nodes. A
dedicated subsystem is offered to manage data staging and replication but data
capabilities are not exposed via pilots. Network resources are assumed to be
available to allow pilots to communicate with the WMS.

Pilots are deployed by TaskQueueDirectors. Three main operations are iterated:
1. getting a list of TaskQueues; 2. calculating the number of pilots to submit
depending on the user-specified priority of each task, and the number and
properties of the available or scheduled pilots; and 3. submitting the
calculated number of pilots.


Natively, DIRAC can execute tasks described by means of the Job Description
Language (JDL)~\cite{pacini2006job}. As such, single-core, multi-core, MPI,
parametric, and collection tasks can be described and submitted. Users can
specify a priority index for each submitted task and one or more specific DCR
that should be targeted for execution. Tasks with complex data dependences can
be described by means of a DIRAC system called ``Transformation Management
System'' (TMS)~\cite{tsaregorodtsev2006dirac}. In this way, user-specified,
data-driven workflows can be automatically submitted and managed by the DIRAC
WMS.

Similar to DIANE and the Coaster System, DIRAC features a task pull model that
naturally implements late binding of tasks to pilots. Each JobWrapper pulls a
new task once it is available and has free resources. No early binding of tasks
on pilots is offered.

\subsubsection{HTCondor Glidein and GlideinWMS}
\label{sec:glidein}

The HTCondor Glidein system~\cite{glidein_manual_url} was developed by the
Center for High Throughput Computing at the University of Wisconsin-Madison
(UW-Madison)~\cite{chtc_url} as part of the HTCondor~\cite{htcondor_url}
software ecosystem. The HTCondor Glidein system implements pilots to aggregate
DCRs with heterogeneous middleware into HTCondor resource pools.

The logical components of HTCondor relevant to the Glidein system are: a set of
Schedd and Startd daemons, a Collector, and a
Negotiator~\cite{glidein_presentation_url}. Schedd is a queuing system that
holds workload tasks and Startd handles the DCR resources. The Collector holds
references to all the active Schedd/Startd daemons, and the Negotiator matches
tasks queued in a Schedd to resources handled by a Startd.

HTCondor Glidein has been complemented by GlideinWMS, a Glidein-based workload
management system that automates deployment and management of Glideins on
multiple types of DCR middleware. GlideinWMS builds upon the HTCondor Glidein
system by adding the following logical components: a set of Glidein Factory
daemons, a set of Frontend daemons for Virtual Organization
(VO)~\cite{foster2001,alfieri2004voms}, and a Collector dedicated to the
WMS~\cite{glideinwms_manual_url}. Glidein Factories submit tasks to the DCRs
middleware, each VO Frontend matches the tasks on one or more Schedd to the
resource attributes advertised by a specific Glidein Factory, and the WMS
Collector holds references to all the active Glidein Factories and VO Frontend
daemons.

\begin{figure}[t]
    \centering
        \includegraphics[width=.48\textwidth]{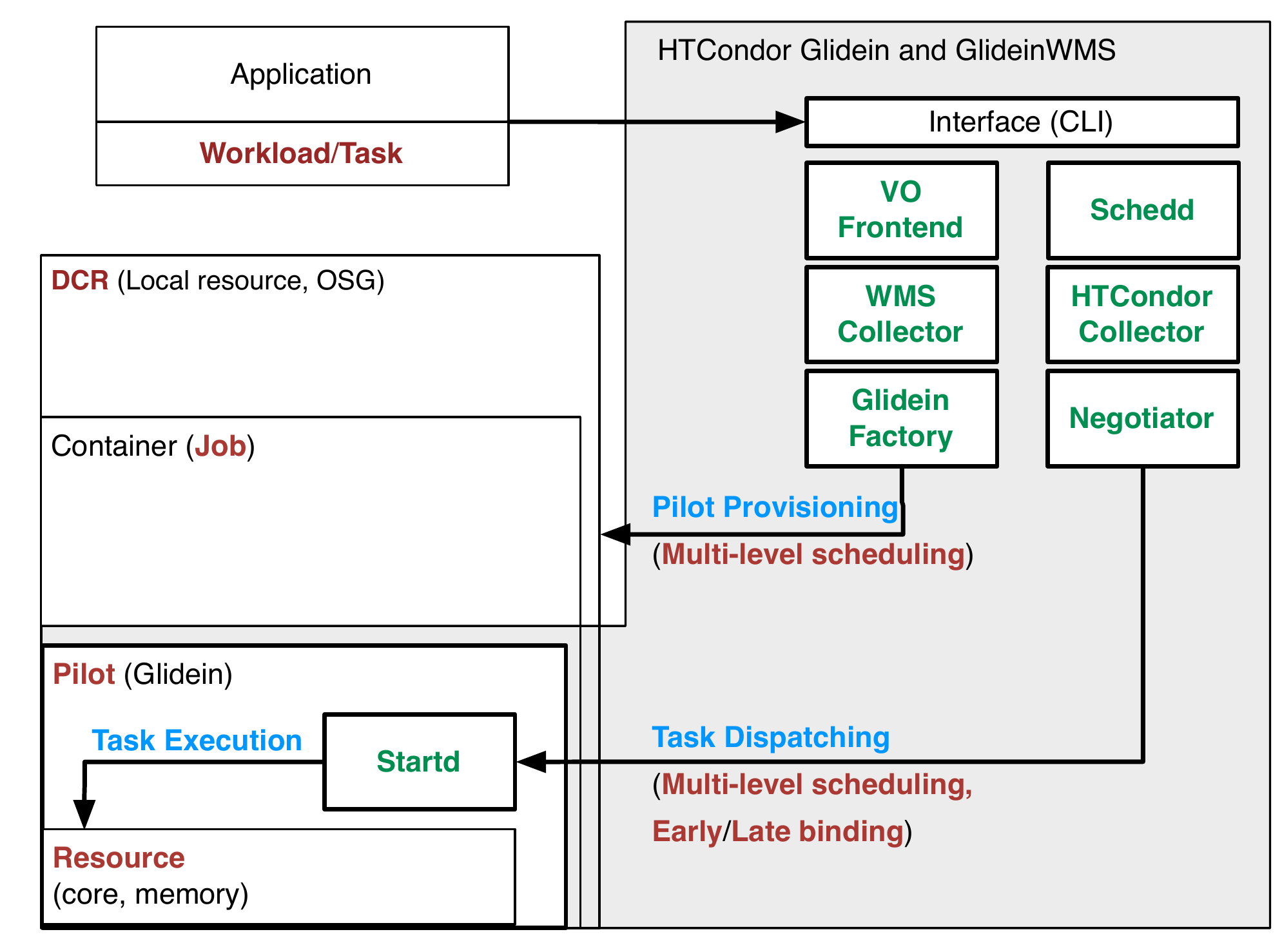}
        \caption{Diagrammatic representation of Glidein components,
          functionalities, and core terminology mapped on
          Figure~\ref{fig:core_terminology}.}
    \label{fig:glidein_comparison}
\end{figure}

Figure~\ref{fig:glidein_comparison} shows the mapping of the HTCondor Glidein
Service and GlideinWMS elements to the components and functionalities of a
\pilot system as described in~\S\ref{sec:understanding}. The set of VO
Frontends and Glidein Factories alongside the WMS collector implement a Pilot
Manager and its pilot provisioning functionality. The set of Schedd, the
Collector, and the Negotiator implement a Workload Manager and its task
dispatching functionality. The Startd daemon implements a Task Manager
alongside its task execution functionality. A Glidein is a job submitted to a
DCR middleware that, once instantiated, configures and executes a Startd
daemon. Glidein is therefore a pilot.

The execution model of the HTCondor Glidein system can be summarized in nine
steps: 1. the user submits a Glidein (i.e., a job) to a DCR batch scheduler; 2.
once executed, this Glidein bootstraps a Startd daemon; 3. the Startd daemon
advertises itself to the Collector; 4. the user submits the tasks of the
workload to the Schedd daemon; 5. the Schedd advertises these tasks to the
Collector; 6. the Negotiator matches the requirements of the tasks to the
properties of one of the available Startd daemon (i.e., a Glidein); 7. the
Negotiator communicates the match to the Schedd; 8. the Schedd submits the
tasks to the Startd daemon indicated by the Negotiator; 9. the task is
executed.

GlideinWMS extends the execution model of the HTCondor Glidein system by
automating the provision of Glideins. The user does not have to submit Glidein
directly but only tasks to Schedd. From there: 1. every Schedd advertises its
tasks with the VO Frontend; 2. the VO Frontend matches the tasks' requirements
to the resource properties advertised by the WMS Connector; 3. the VO Frontend
places requests for Glideins instantiation to the WMS Collector; 4. the WMS
Collector contacts the appropriate Glidein Factory to execute the requested
Glideins; 5. the requested Glideins become active on the DCRs; and 6. the
Glideins advertise their availability to the (HTCondor) Collector. From there
on the execution model is the same as described for the HTCondor Glidein
Service.


The resources managed by a single Glidein (i.e., pilot) are limited to compute
resources. Glideins may bind one or more cores, depending on the target DCRs.
For example, heterogeneous HTCondor pools with resources for desktops,
workstations, small campus clusters, and some larger clusters will run mostly
single core Glideins. More specialized pools that hold, for example, only DCRs
with HTC, Grid, or Cloud middleware may instantiate Glideins with a larger
number of cores. Both HTCondor Glidein and GlideinWMS provide abstractions for
file staging but pilots are not used to hold data or network resources.

The process of pilot deployment is the main difference between HTCondor Glidein
and GlideinWMS. While the HTCondor Glidein system requires users to submit the
pilots to the DCRs, GlideinWMS automates and optimizes pilot provisioning.
GlideinWMS attempts to maximize the throughput of task execution by
continuously instantiating Glideins until the queues of the available Schedd
are emptied. Once all the tasks have been executed, the remaining Glideins are
terminated.


HTCondor Glidein and GlideWMS expose the interfaces of HTCondor to the
application layer and no theoretical limitation is posed on the type and
complexity of the workloads that can be executed. For example, DAGMan (Directed
Acyclic Graph Manager)~\cite{couvares2007workflow} has been designed to execute
workflows by submitting tasks to Schedd, and a tool is available to design
applications based on the \MW coordination pattern.



HTCondor was originally designed for resource scavenging and opportunistic
computing. Thus, in practice, independent and single (or few-core) tasks are
more commonly executed than many-core tasks, as is the case for OSG, the
largest HTCondor and GlideinWMS deployment. Nonetheless, in principle projects
may use dedicated installation and resources to execute tasks with larger core
requirements both for distributed and parallel applications, including MPI
applications.

Both HTCondor Glidein and GlideWMS rely on one or more HTCondor Collectors to
match task requirements and resource properties, represented as
ClassAds~\cite{classad_url}. This matching can be evaluated right before the
scheduling of the task. In this way, late binding is achieved but early binding
remains unsupported.

\subsubsection{MyCluster}\label{sec:mycluster}

MyCluster~\cite{walker2007personal,mycluster_url} is not maintained but is
included in the comparison because it presents some distinctive features. Its
user/\pilot system interface and task submission system based on the notion of
virtual cluster highlight the flexibility of \pilot systems implementations.
Moreover, MyCluyster was one of the first \pilot system to be aimed
specifically at HPC DCRs.


MyCluster was originally developed at the Texas Advanced Computing Center
(TACC)~\cite{tacc_url}, sponsored by NSF to enable execution of workloads on
TeraGrid~\cite{teragrid_url}, a set of DCRs deploying Grid middleware.
MyCluster provides users with virtual clusters: aggregates of homogeneous
resources dynamically acquired on multiple and diverse DCRs. Each virtual
cluster exposes HTCondor, SGE~\cite{chase2003dynamic}, or
OpenPBS~\cite{openpbs_url} job-submission systems, depending on the user and
use case requirements.

MyCluster is designed around three main components: a Cluster Builder Agent, a
system where users create Virtual Login Sessions, and a set of Task Managers.
The Cluster Builder Agent acquires the resources from diverse DCRs by means of
multiple Task Managers, while the Virtual Login Session presents these
resources as a virtual cluster to the user. A virtual login session can be
dedicated to a single user, or customized and shared by all the users of a
project. Upon login on the virtual cluster, a user is presented with a
shell-like environment used to submit tasks for execution.

\begin{figure}[t]
    \centering
        \includegraphics[width=.48\textwidth]{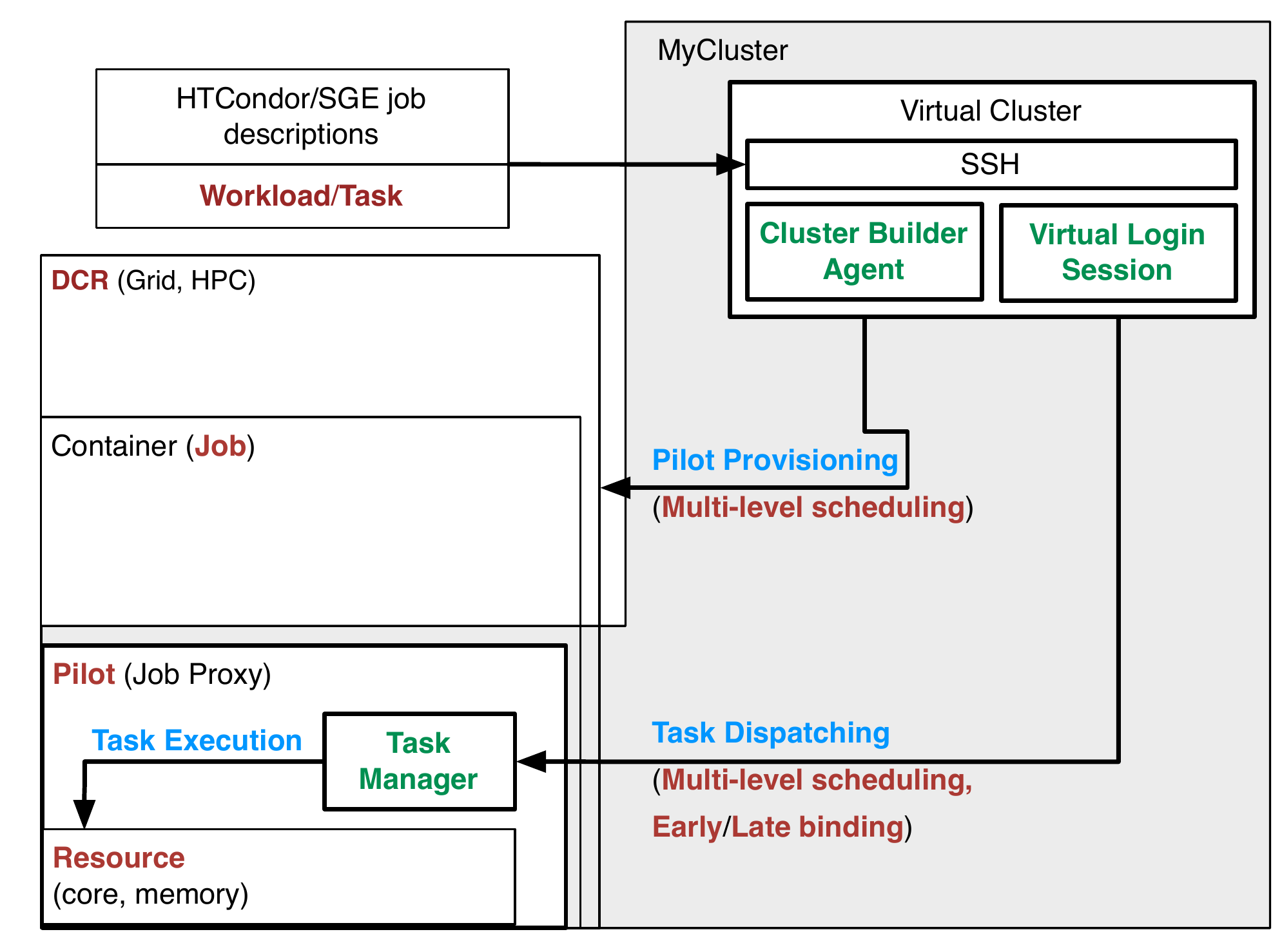}
    \caption{Diagrammatic representation of MyCluster components,
    functionalities, and core terminology mapped on
    Figure~\ref{fig:core_terminology}.}
    \label{fig:mycluster_comparison}
\end{figure}

Figure~\ref{fig:mycluster_comparison} shows how the components of MyCluster map
to the components and functionalities of a \pilot system as described
in~\S\ref{sec:compsandfuncs}: The Cluster Builder Agent implements a Pilot
Manager and a Virtual Login Session implements a Workload Manager. The Task
Manager shares its name and functionality with the homonymous component defined
in~\S\ref{sec:compsandfuncs}. The Cluster Builder Agent provides Task Managers
by submitting Job Proxies to diverse DCRs, and a Virtual Login Session uses the
Task Managers to submit and execute tasks. As such, Job Proxies are pilots.

The execution model of MyCluster can be summarized in five steps: 1. a user
logs into a dedicated virtual cluster via, for example, ssh to access a
dedicated Virtual Login Session; 2. the user writes a job wrapper script using
the HTCondor, SGI, or OpenPBS job specification language; 3. the user submits
the job to the job submission system on the virtual cluster; 4. the Cluster
Builder Agent submits a suitable number of Job Proxies on one or more DCR; 5.
when the Job Proxies become active, the user-submitted job is executed on the
resources they hold.

Job Proxies hold compute resources in the form of compute cores. MyCluster does
not offer any dedicated data subsystem and Job Proxies (i.e. pilots) are not
used to expose data resources to the user. Users are assumed to stage the data
required by the compute tasks directly, or by means of the data capabilities
exposed by the job submission system of the virtual cluster. Networking is
assumed to be available among the MyCluster components.

The Cluster Builder Agent submits Job Proxies to each DCR by using the
GridShell framework~\cite{walker2004gridshell}. GridShell wraps the Job Proxies
description into the job description language supported by the target DCR.
Thanks to GridShell, MyCluster can submit jobs to DCR with diverse middleware.

MyCluster exposes a virtual cluster with a predefined job submission system to
the user. Pilots can have a user-defined amount of cores inter or cross-compute
node. As such, every application built to utilize HTCondor, SGE, or OpenPBS can
be executed transparently on MyCluster. This includes single and multi-core
tasks, MPI tasks, and data-driven workflows.

The jobs specified by a user are bound to the DCR resources as soon as Job
Proxies become active. The user does not have to specify on which Job Proxies
or DCR each task has to be executed. In this way, MyCluster implements late
binding.

%
\subsubsection{PANDA}
\label{sec:panda}

\panda (Production and Distributed Analysis)~\cite{zhao2011panda} was developed
to provide a workload management system (WMS) for ATLAS. ATLAS is a particle
detector at the LHC that requires a WMS to handle large numbers of tasks for
their data-driven processing workloads. In addition to the logistics of
handling large-scale task execution, ATLAS also needs integrated monitoring for
the analysis of system state, and a high degree of automation to reduce user
and administrative intervention.

\panda has been initially deployed as a HTC-oriented, multi-user WMS system
consisting of ~100 heterogeneous computing sites~\cite{maeno2012pd2p}. Recent
improvements to \panda have extended the range of deployment scenarios to HPC
and cloud-based DCRs making \panda a general-purpose \pilot
system~\cite{nilsson2012recentrp}.

\panda architecture consists of a Grid Scheduler and a \panda
Server~\cite{panda_architecture_url,maeno2011overview}. The Grid Scheduler is
implemented by a component called ``AutoPilot'' that submits jobs to diverse
DCRs. The \panda server is implemented by four main components: a Task Buffer,
a Broker, a Job Dispatcher, and a Data Service. The Task Buffer collects all
the submitted tasks into a global queue and the Broker prioritizes and binds
those tasks to DCRs on the basis of multiple criteria. The Data Service stages
the input file(s) of the tasks to the DCR to which the tasks have been bound
using the data transfer technologies exposed by the DCR middleware (e.g.,
uberftp, gridftp, or lcg-cp). The Job Dispatcher delivers the tasks to the
RunJobs run by each Pilot bound to a DCR.

\begin{figure}[t]
    \centering
        \includegraphics[width=.48\textwidth]{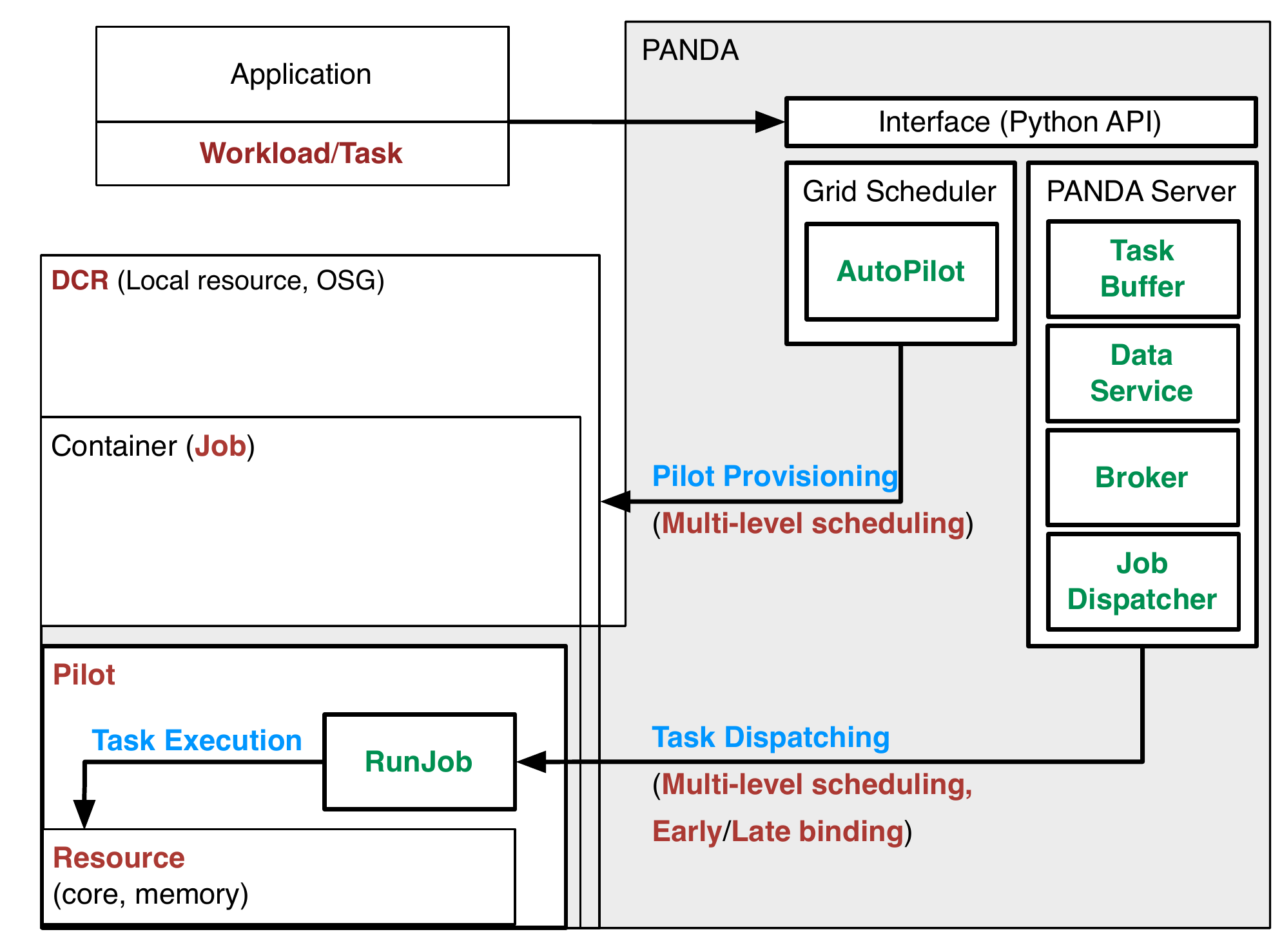}
    \caption{Diagrammatic representation of PANDA components, functionalities,
    and core terminology mapped on Figure~\ref{fig:core_terminology}.}
    \label{fig:panda_comparison}
\end{figure}

Figure~\ref{fig:panda_comparison} shows how PANDA implements the components and
functionalities of a \pilot system as described in~\S\ref{sec:understanding}:
the Grid Scheduler is a Pilot Manager implementing Pilot Provisioning while the
\panda Server is a Workload Manager implementing Task Dispatching. The jobs
submitted by the Grid Scheduler are called ``Pilots'' and act as pilots once
instantiated on the DCR by running RunJob, i.e., the Task Manager. RunJob
contacts the Job Dispatcher component to request for tasks to be executed.

The execution model of PANDA can be summarized in eight
steps~\cite{nilsson2011atlas,pandarun_url}: 1. the user submits tasks to the
\panda server; 2. the tasks are queued within the Task Buffer; 3. the tasks
requirements are evaluated by the Broker and bound to a DCR; 4. the input files
of the tasks are staged to the bound DCR by the Data Service; 5. the required
pilot(s) are submitted as jobs to the target DCR; 6. the submitted pilot(s)
becomes available and reports back to the Job Dispatcher; 7. tasks are
dispatched to the available pilots for execution; 8. tasks are executed.


\panda pilots expose mainly single cores, but extensions have been developed to
instantiate pilots with multiple cores~\cite{crooks2012multi}. The Data Service
of \panda allows the integration and automation of data staging within the task
execution process, but no pilots are offered for data~\cite{maeno2012pd2p}.
Network resources are assumed to be available among \panda components, but no
network-specific abstraction is made available.

The AutoPilot component of \panda's Grid Scheduler has been designed to use
multiple methods to submit pilots to DCRs. The \panda installations of the US
ATLAS infrastructure uses the HTCondor-G~\cite{frey2002condorG} system to
submit pilots to the US production sites. Other schedulers enable AutoPilot to
submit to local and remote batch systems and to the GlideinWMS frontend.
Submissions via the canonical tools offered by HTCondor have also been used to
submit tasks to cloud resources.

\panda was initially designed to serve specifically the ATLAS use case,
executing mostly single-core tasks with input and output files. Since its
initial design, the ATLAS analysis and simulation tools have started to
investigate multi-core task execution with AthenaMP~\cite{crooks2012multi} and
\panda has been evolving towards a more general purpose workload
manager~\cite{schovancova2014next,schovancova2013panda,borodin2015scaling}.
Currently, \panda offers experimental support for multi-core pilots and tasks
with or without data dependences. \panda is being generalized to support
applications from a variety of science domains.\cite{maeno2014evolution}.

\panda offers late binding but not early binding capabilities. Workload jobs
are assigned to activated and validated pilots via the \panda server based on
brokerage criteria like data locality and resource characteristics.

%
\subsubsection{RADICAL-Pilot}
\label{sec:radical_pilot}

The authors of this paper have been engaged in theoretical and practical
aspects of \pilot systems. In addition to formulating the P*
Model~\cite{pstar12}, the RADICAL group~\cite{radical_url} is responsible for
developing and maintaining the RADICAL-Pilot \pilot system~\cite{rp_url}.
RADICAL-Pilot is built upon the experience gained from developing BigJob, and
integrating it with many
applications~\cite{ko-efficient,ecmls_ccpe10,ct500776j} on different DCRs.

RADICAL-Pilot consists of five main logical components: a Pilot Manager, a
Compute Unit (CU) Manager, a set of Agents, the SAGA-Python DCR interface, and
a database. The Pilot Manager describes pilots and submits them via SAGA-Python
to DCR(s), while the CU manager describes tasks (i.e. CU) and schedules them to
one or more pilots. Agents are instantiated on DCRs and execute the CUs pushed
by the CU manager. The database is used for the communication and coordination
of the other four components.

\begin{figure}[t]
    \centering
        \includegraphics[width=.48\textwidth]{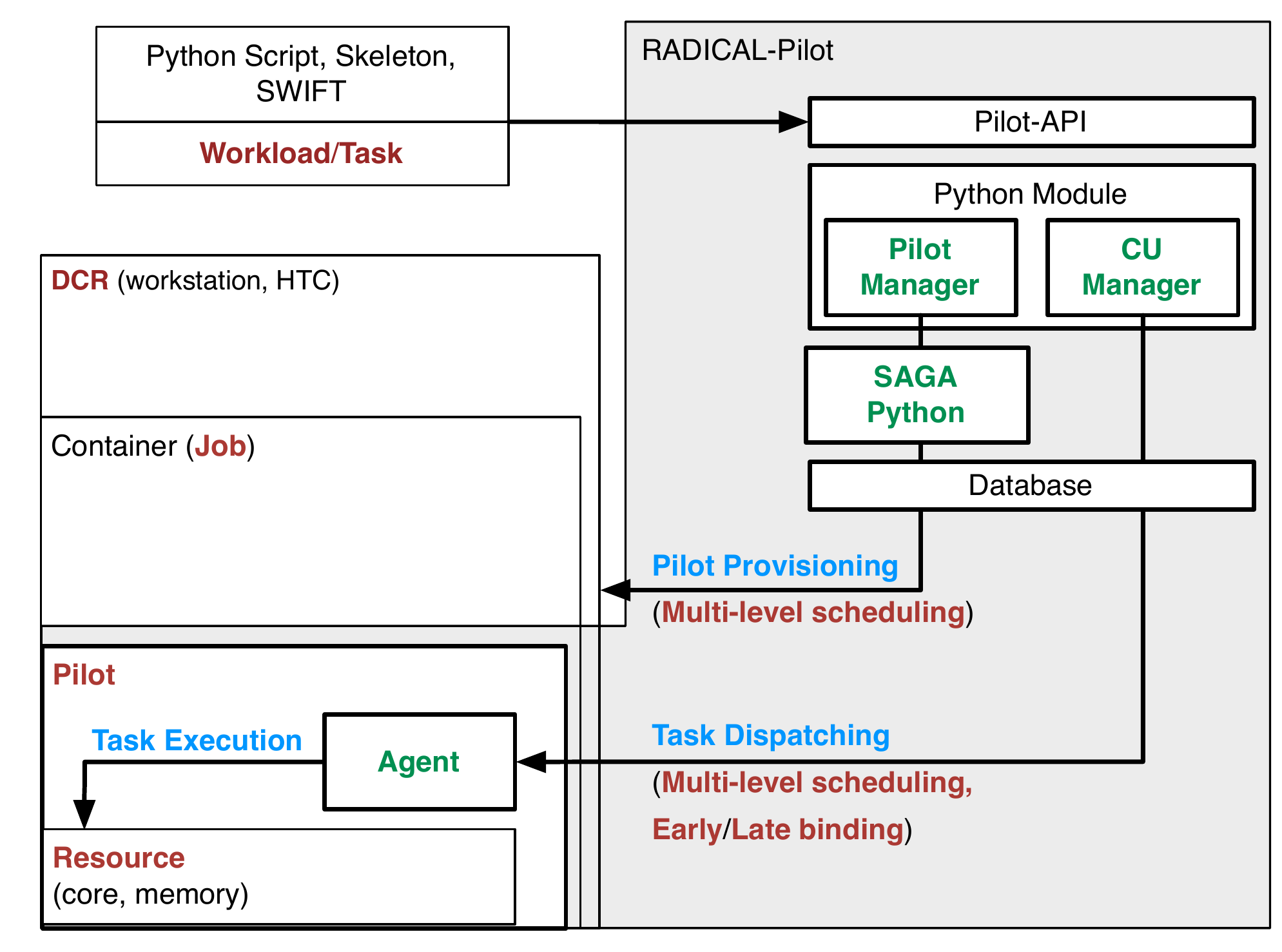}
    \caption{Diagrammatic representation of RADICAL Pilot components,
    functionalities, and core terminology mapped on
    Figure~\ref{fig:core_terminology}.}
    \label{fig:rp_comparison}
\end{figure}

RADICAL-Pilot closely resembles the description offered
in~\S\ref{sec:understanding} (see Figure~\ref{fig:rp_comparison}). The Pilot
Manager and SAGA-Python implement the logical component also called ``Pilot
Manager'' in~\S\ref{sec:compsandfuncs}. The Workload Manager is implemented by
the CU Manager. The Agent is deployed on the DCR to expose its resources and
execute the tasks pushed by the CU Manager. As such, the Agent is a pilot.


RADICAL-pilot is implemented as two Python modules to support the development
of distributed applications. The execution model of RADICAL-Pilot can be
summarized in six steps: 1. the user describes tasks in Python as a set of CUs
with or without data and DCR dependences; 2. the user also describes one or
more pilots choosing the DCR(s) they should be submitted to; 3. upon
execution of the user's application, the Pilot Manager submits each pilot that
has been described to the indicated DCR utilizing the SAGA interface; 4. The CU
Manager schedules each CU either to the pilot indicated in the CU or on the
first pilot with free and available resources. Scheduling is done by storing
the CU description into the database; 5. when required, the CU Manager also
stages the CU's input file(s) to the target DCR; and 6. the Agent pulls its CU
from the database and executes it.

The Agent component of RADICAL-Pilot offers abstractions for both compute and
data resources. Every Agent can expose between one and all the cores of the
compute node where it is executed; it can also expose a data handle that
abstracts away specific storage properties and capabilities. In this way, the
CUs running on an Agent can benefit from unified interfaces to both core and
data resources. Networking is assumed to be available between the RADICAL-Pilot
components.

The Pilot Manager deploys the Agents of RADICAL-Pilot by means of the
SAGA-python API~\cite{saga-x}. SAGA provides access to diverse DCR middleware
via a unified and coherent API, and thus RADICAL-Pilot can submit pilots to
resources exposed by XSEDE and NERSC~\cite{nersc_url}, by the OSG HTCondor
pools, and many ``leadership'' class systems like those managed by
OLCF~\cite{olcf_url} or NCSA~\cite{ncsa_url}.

The resulting separation of agent deployment from DCR architecture reduces the
overheads of adding support for a new DCR~\cite{review_radicalpilot_2015}. This
is illustrated by the relative ease with which RADICAL-Pilot is extended to
support (i) a new type of DCR such as IaaS, and (ii) DCRs that have essentially
similar architecture but different middleware, for example the Cray
supercomputers operated in the US and Europe.

RADICAL-Pilot can execute tasks with varying coupling and communication
requirements. Tasks can be completely independent, single or multi-threaded;
they may be loosely coupled requiring input and output files dependencies, or
they might require low-latency runtime communication. As such, RADICAL-Pilot
supports MPI applications, workflows, and diverse execution patterns such as
pipelines.

CU descriptions may or may not contain a reference to the pilot to which the
user wants to bind the CU. When a reference is present, the scheduler of the CU
Manager waits for a slot to be available on the indicated pilot. When a target
pilot is not specified, the CU Manager binds and schedules the CU on the first
pilot available. As such, RADICAL-Pilot supports both early and late binding,
depending on the use case and the user specifications.

%
\subsection{Comparison}
\label{sec:context}

The previous subsection shows how diverse \pilot system implementations conform
to the architecture pattern we described in~\S\ref{sec:termsdefs}. This
confirms the generality of the pattern at capturing the components and
functionalities required to implement a \pilot system. The described \pilot
systems also show implementation differences, especially concerning the
following auxiliary properties: Architecture, Communication and Coordination,
Interoperability, Interface, Security, and Performance and Scalability.

The \pilot systems described in~\S\ref{sec:implementations} implement different
architectures. DIANE, DIRAC, and, to some extent, both PANDA and the Coaster
System are monolithic (Figures~\ref{fig:diane_comparison},
\ref{fig:dirac_comparison}, \ref{fig:panda_comparison}, and
\ref{fig:coaster_comparison}). Most of their functionalities are aggregated
into a single component implemented ``as a service''~\cite{erl2005service}. A
dedicated hardware infrastructure is assumed for a production-grade deployment
of DIRAC and PANDA. Consistent with a Globus-oriented design, the Coaster
Service is instead assumed to be run on the DCR acting as a proxy for both the
pilot and workload functionalities.

MyCluster and RADICAL-Pilot also are mostly monolithic
(Figures~\ref{fig:rp_comparison} and~\ref{fig:mycluster_comparison}) but not
implemented as a service. MyCluster resembles the architecture of a HPC
middleware while Radical-Pilot is implemented as two Python modules. MyCluster
requires dedicated hardware analogously to the head-node of a traditional HPC
cluster. RADICAL-Pilot users are instead free to decide where to deploy their
applications, either locally on workstations or remotely on dedicated machines.
In production-grade deployment, RADICAL-Pilot requires a dedicated database to
support its communication and coordination protocols.


GlideinWMS requires integration within the HTCondor ecosystem and therefore
also a service oriented architecture but it departs from a monolithic design.
GlideinWMS implements a set of separate, mostly autonomous services
(Figure~\ref{fig:glidein_comparison}) that can be deployed depending on the
available resources and on the motivating use case.

Architecture frameworks and description
languages~\cite{medvidovic2000classification,al_url} can be used to further
specify and refine the component architectures in
Figures~\ref{fig:coaster_comparison}-\ref{fig:rp_comparison}. For example, the
4+1 framework alongside a UML-based
notation~\cite{uml_url,medvidovic1999assessing} could be used to describe
multiple ``views'' of each \pilot system architecture, offering more details
and better documentation about the implementation of their components, the
functionalities provided to the user, the behavior of the system, and its
deployment scenarios.

The \pilot systems described in the previous subsection also display
differences in their communication and coordination models. While all the
\pilot systems assume preexisting networking functionalities, the Coaster
System implements a dedicated communication protocol used both for coordination
and data staging. The Coaster System and RADICAL-Pilot can both work as
communication proxies among the \pilot system's components when the DCR compute
nodes do not expose a public network interface. All the \pilot systems
implement the \MW coordination pattern, but the Task and the Workload Managers
in DIRAC, PANDA, MyCluster, and the Coaster System can also coordinate to
recover task failures and isolate under-performing or failing DCR compute
nodes.

Figures~\ref{fig:coaster_comparison}-\ref{fig:rp_comparison} also shows
different interfaces between \pilot systems and DCRs, and between \pilot
systems and users or applications. Most of the described \pilot systems
inter-operate across diverse DCR middleware, including HPC, grid, and cloud
batch systems. Implementations of this interoperability diverge, ranging from
the dedicated SAGA API used by RADICAL-Pilot, to special-purpose connectors
used by DIANE, DIRAC and PANDA, to the installation of specialized components
on the DCR middleware used by Coaster System, Glidein, and MyCluster. These
interfaces are functionally analogous; reducing their heterogeneity would limit
effort duplication and promote interoperability across \pilot systems.

The interfaces exposed to give users access to pilot capabilities differ both
in types and implementations. DIANE, DIRAC, GlideinWMS, MyCluster, and PANDA
offer command line tools. These are often tailored to specific use cases,
applications, and DCRs, requiring to be installed on the users' workstations or
on dedicated machines. The Coaster System and RADICAL-Pilot expose an API, and
the command line tools of DIANE, DIRAC, and PANDA are built on APIs that users
may directly access to develop distributed applications.

Differences in the user interfaces stem from assumptions about distributed
applications and their use cases. Interfaces based on command line tools assume
applications that can be ``submitted'' to the \pilot system for execution. APIs
assume instead applications that need to be coded by the user, depending on the
specific requirements of the use case. These assumptions justify multiple
aspects of the design of \pilot systems, determining many characteristics of
their implementations.

The described \pilot systems also implement different types of authentication
and authorization (AA). The AA required by the user to submit tasks to their
own pilots varies depending on the pilot's tenancy. With single tenancy, AA can
be based on inherited privileges as the pilot can be accessed only by the user
that submitted it. With multitenancy, the \pilot system has to evaluate whether
a user requesting access to a pilot is part of the group of allowed users. This
requires abstractions like virtual organizations and certificate
authorities~\cite{horwitz2002}, implemented, for example, by GlideinWMS and the
Coaster Systems.

The credential used for pilot deployment depends on the target DCR. The AA
requirements of DCRs are a diverse and often inconsistent array of mechanisms
and policies. \pilot systems are gregarious in the face of such a diversity as
they need to present the credentials provided by the application layer (or
directly by the user) to the DCR. As such, the AA requirements specific to
\pilot systems are minimal but the implementation required to present suitable
credentials may be complex, especially when considering
\pilot systems offering interoperability among diverse DCRs.

Finally, the differences among \pilot system implementations underline the
difficulties in defining and correlating performance metrics. The performance
of each \pilot system can be evaluated under multiple metrics that are affected
by the workload, the \pilot system behavior, and the DCR. For example, the
commonly used metrics of system overhead and workload's time to completion
depend on the design of the \pilot system; on the data, compute and network
requirements of the workload executed; and on the capabilities of the target
resources. These parameters vary at every execution and require dedicated
instrumentation built into the \pilot system to be measured. Without consistent
performance models and set of probes, performance comparison among \pilot
systems appears unfeasible.

%
\section{Discussion and Conclusion}
\label{sec:discussion}

We introduced the \pilot abstraction in \S\ref{sec:understanding} describing
the capabilities, components, and architecture pattern of \pilot systems. We
also defined  a terminology consistent across \pilot systems clarifying the
meaning of ``pilot'', ``job'', and their cognate concepts. In
\S\ref{sec:analysis} we offered a classification of the core and auxiliary
properties of \pilot system implementations, and we analyzed a set of exemplars.
Considered altogether, these contributions outline a paradigm for the
execution of heterogeneous, multi-task workloads via multi-entity and
multi-stage scheduling on DCR resource placeholders. This computing paradigm is
here referred to as ``\pilot paradigm''.

%
\subsection{The Pilot Paradigm}
\label{sec:paradigm}

The generality of the \pilot paradigm may come as a surprise when considering
that, traditionally, \pilot systems have been implemented to optimize the
throughput of single-core (or at least single-node), short-lived, uncoupled
tasks execution~\cite{pordes2007,sfiligoi2009,juve2010}. For example DIANE,
DIRAC, MyCluster, \panda, or HTCondor Glidein and GlideinWMS were initially
developed to focus on either a type of workload, a specific infrastructure, or
the optimization of a single performance metric.

The \pilot paradigm is general because the execution of a workload via
multi-entity and multi-stage scheduling on DCR resource placeholders does not
have to depend on a single type of workload, DCR, or resource. In principle,
systems implementing the \pilot paradigm can execute workloads composed of an
arbitrary number of tasks with diverse data, compute, and networking
requirements. The same generality applies to the types of DCR and of resource
on which a \pilot system executes workloads.\footnote{The generality of the
pilot paradigm across workload, DCR, and resource types was first discussed in
Ref.~\cite{pstar12}, wherein an initial conceptual model for \pilot systems was
proposed. The introduction of the pilot architecture pattern and the discussion
in~\S\ref{sec:understanding} and~\S\ref{sec:analysis} enhances and extends the
preliminary analysis of Ref.~\cite{pstar12}.}

The analysis presented in~\S\ref{sec:analysis}, shows how \pilot systems have
progressed to implement the generality of the \pilot paradigm. \pilot systems
are now engineered to execute homogeneous or heterogeneous workloads; these
workloads can be comprised of independent or intercommunicating tasks of
arbitrary duration or data and computation requirements. These workloads can
also be executed on an increasingly diverse pool of DCRs. \pilot systems were
originally designed for DCR with HTC grid middleware; \pilot systems have
emerged that are capable of also operating on DCRs with HPC and cloud
middleware.

As seen in~\S\ref{sec:understanding}, the \pilot paradigm demands resource
placeholders but does not specify the type of resource that the placeholder
should expose. In principle, pilots can also be placeholders for data or
network resources, either exclusively or in conjunction with compute resources.
For example, in Ref.~\cite{pilot-data-jpdc-2014} the concept of \pilotdata was
conceived to be fundamental to dynamic data placement and scheduling as \pilot
is to computational tasks. The concept of ``Pilot networks'' was introduced in
Ref.~\cite{pstar-network12} in reference to Software-Defined
Networking~\cite{kirkpatrick2013software} and User-Schedulable Network
paths.\cite{he2013software}

The generality of the \pilot paradigm also promotes the adoption of \pilot
functionalities and systems by other middleware and tools.
For example, \pilot systems have been
successfully integrated within workflow systems to support optimal execution of
workloads with articulated data and single or multi-core task
dependencies~\cite{zhao2007swift,camarasu2010dynamic,deelman2015}. As such, not
only can throughput be optimized for multi-core, long-lived, coupled tasks
executions, but also for optimal data/compute placement, and dynamic resource
sizing.

The \pilot paradigm is not limited to academic projects and scientific
experiments. Hadoop~\cite{hadoop_url} introduced the
YARN~\cite{vavilapalli2013apache} resource manager for heterogeneous workloads.
YARN supports multi-entity and multi-stage scheduling: applications initialize
an ``Application-Master'' via YARN; the Application Master allocates resources
in ``containers'' for the applications; and YARN then can execute tasks in
these containers (i.e., resource placeholders). TEZ~\cite{tez_url}, a DAG
processing engine primarily designed to support the Hive SQL
engine~\cite{thusoo2009hive}, enables applications to hold containers across
the DAG execution without de/reallocating resources. Independent of the Hadoop
developments, Google's Kubernetes~\cite{bernstein2014containers} is emerging as
a leading container management approach. Not completely coincidently,
Kubernetes is the Greek term for the English ``Pilot''.

%
\subsection{Future Directions and Challenges}
\label{sec:future}

The \pilot landscape is currently fragmented with duplicated effort and
capabilities. The reasons for this balkanization can be traced back mainly to
two factors: (i) the relatively recent discovery of the generality and
relevance of the \pilot paradigm; and (ii) the development model fostered
within academic institutions.

As seen in~\S\ref{sec:history} and~\S\ref{sec:analysis}, \pilot systems were
developed as a pragmatic solution to improve the throughput of distributed
applications, and were designed as local and point solutions. \pilot systems
were not thought from their inception as an independent system, but, at best,
as a module within a framework. Inheriting the development model of the
scientific projects within which they were initially developed, \pilot systems
were not engineered to promote (re)usability, modularity, open interfaces, or
long-term sustainability. Collectively, this resulted in duplication of
development effort across frameworks and projects, and hindered the
appreciation for the generality of the \pilot abstraction, the theoretical
framework underlying the \pilot systems, and the paradigm for application
execution they enable.

Consistent with this analysis, many of the \pilot systems described
in~\S\ref{sec:implementations} offer a set of overlapping functionalities. This
duplication may have to be reduced in the future to promote maintainability,
robustness, interoperability, extensibility, and overall capabilities of
existing \pilot systems. As seen in~\S\ref{sec:context}, \pilot systems are
already progressively supporting diverse DCRs and types of workload. This trend
might lead to consolidation and to increased adoption of multi-purpose
\pilot systems. The scope of the consolidation process will depend on the
diversity of used programming languages, deployment models, interaction with
existing applications, and how they will be addressed.

The analysis proposed in this paper suggests critical commonalities across
\pilot systems stemming from a shared architectural pattern, abstraction, and
computing paradigm. Models of pilot functionality can be grounded on these
commonalities, as well as be reflected in the definition of unified and open
interfaces for the users, applications, and DCRs. End-users, developers, and
DCR administrators could rely upon these interfaces, which would promote better
integration of \pilot systems into application and resource-facing middlware.


There is evidence of ongoing integration and consolidation processes, such as
the adoption of extensible workload management capabilities or utilization of
similar resource interoperability layers.  For example, \panda is iterating its
development cycle and the resulting system, called ``Big \panda'' is now
capable of opportunistically submitting pilots to the Titan
supercomputer~\cite{titan_url} at the Oak Ridge Leadership Computing Facility
(OLCF)~\cite{panitkin2015integration,olcf_url}.  Further, Big \panda has
adopted SAGA, an open and standardized DCR interoperability library developed
independent of \pilot systems but now adopted both by Big Panda and
RADICAL-Pilot.

%
\subsection{Summary and Contributions}
\label{sec:contributions}

This paper contributes to the understanding, design, and adoption of \pilot
systems by characterizing the \pilot abstraction, the
\pilot paradigm, and exemplar implementations.

We provided an analysis of the technical origins and motivations of \pilot
systems in~\S\ref{sec:history} and we summarized their chronological development
in Figure~\ref{fig:timeline}. We described the logical components and
functionalities that constitute the \pilot abstraction
in~\S\ref{sec:understanding}, and we outlined them in
Figure~\ref{fig:comp_func}.  We then defined a consistent terminology to clarify
the heterogeneity of the \pilot systems landscape, and we used this terminology
together with the logical components and functionalities of the \pilot systems
to specify the pilot architecture pattern in Figure~\ref{fig:core_terminology}.

We defined the core and auxiliary properties of \pilot system implementations
in~\S\ref{sec:analysis} (Tables~\ref{table:core_properties}
and~\ref{table:aux_properties}). We then used these properties alongside the
contributions offered in~\S\ref{sec:understanding} to describe seven exemplar
\pilot system implementations. We gave details about their architecture and
execution model showing how they conformed to the pilot architecture paradigm
we defined in~\S\ref{sec:termsdefs}. 
We summarized this analysis in
Figures~\ref{fig:coaster_comparison}--\ref{fig:rp_comparison}.

We used the \pilot abstraction and insight about \pilot systems, their
motivations and diverse implementations to highlight the properties of the
\pilot paradigm in~\S\ref{sec:discussion}.  We argued for the generality of the
\pilot paradigm on the basis of demonstrated generality of the type of workload
and use cases \pilot systems can execute, as well as a lack of constraints on
the type of DCR that can be used or on the type of resource exposed by the
pilots.  Finally, we reviewed the benefits that a more structured approach to
the conceptualization and design of \pilot systems may offer.



With this paper, we also contributed a methodology to evaluate software systems
that have developed organically and without an established theoretical
framework. This methodology is composed of five steps: (i) analysis of the
abstraction(s) underlying the observed software system implementations; (ii)
the definition of a consistent terminology to reason about abstractions; (iii)
the evaluation of the components and functionalities that may constitute a
specific architectural pattern for the implementation of that abstraction; (iv)
the definition of core and auxiliary implementation properties; (v) the
evaluation of implementations.

The application of this methodology offers the opportunity to uncover the
theoretical framework underlying the observed software systems, and to
understand whether such systems are implementations of a well-defined and
independent abstraction. This theoretical framework can be used to inform or
understand the development and engineering of software systems without mandating
specific design, representation, or development methodologies or tools.

Workflow systems are amenable to be studied with the methodology proposed and
used in this paper.  Multiple workflow systems have been developed
independently to serve diverse use cases and be executed on heterogeneous DCRs.
In spite of broad surveys~\cite{taylor2007workflows,yu2005taxonomy,barker2008scientific,blythe2005task} about workflow systems and their usage
scenarios, an encompassing theoretical framework for the underlying
abstraction, or set of abstractions if any, is not yet available. This is
evident in the state of workflow systems which shows a significant duplication
of effort, limited extensibility and interoperability, and proprietary
solutions for interfaces to both the resource and application layers.

%
\section*{Acknowledgements and Author Contributions}

This work is funded by the Department of Energy Award (ASCR) DE-FG02-12ER26115,
NSF CAREER ACI-1253644 and NSF ACI-1440677 ``RADICAL-Cybertools''. We thank
former and current members of the RADICAL group for helpful discussions,
comments, and criticisms. We also thank members of the AIMES project for
helpful discussions, in particular Daniel S. Katz for comments. We thank Rajiv
Ramnath for useful discussions that helped to establish connections between
this work and software architecture patterns.  MT wrote the paper. MT and SJ
co-organized and edited the paper. MS contributed to an early draft of parts of
Section 2 and 4.

%
\bibliographystyle{IEEEtran}
\bibliography{pilotreview,radical_publications}

\end{document}